\documentclass[12pt]{article}
\usepackage{graphicx}
\usepackage{mathptmx,mathrsfs}
\usepackage{amsfonts,amsmath,amssymb,amsthm}
\usepackage{indentfirst}
\usepackage{cite}
\usepackage{hyperref}
\usepackage{enumitem}

\numberwithin{equation}{section}
\theoremstyle{definition}

\begin{document}

\title{The Epstein-Glaser causal approach to the Light-Front QED$_{4}$. I: Free theory}
\author{R. Bufalo$^{1}$\thanks{%
rbufalo@ift.unesp.br}~, B.M. Pimentel$^{1}$\thanks{%
pimentel@ift.unesp.br}~, D.E. Soto$^{1}$\thanks{%
danielsb@ift.unesp.br}~ \\
\textit{{\small $^{1}$ Instituto de F\'{\i}sica Te\'orica (IFT), UNESP, S\~ao Paulo State University} }\\
\textit{ {\small Rua Dr. Bento Teobaldo Ferraz 271, Bloco II Barra Funda, CEP
01140-070 S\~ao Paulo, SP, Brazil}}\\
}
\maketitle
\date{}

\begin{abstract}
In this work we present the study of light-front field theories in the realm of the axiomatic theory.
It is known that when one uses the light-cone gauge pathological poles $\left( k^{+}\right) ^{-n}$ arises,
demanding a prescription to be employed in order to tame these ill-defined poles and to have the correct Feynman integrals due to the lack of Wick rotation in such theories. In order to shed a new light on this long standing problem we
present here a discussion based on the use of rigorous mathematical machinery of the distributional theory combined with physical concepts, such as causality, to show how to deal with these singular propagators in a general fashion without making use of any prescription. The first step of our development will consist in showing how the analytic representation for propagators arises by requiring general physical properties within the framework of Wightman's formalism. From that we shall determine the equal-time (anti)commutation relations in the light-front form for the scalar and fermionic fields, as well as for the dynamical components of the electromagnetic field. In conclusion, we introduce the Epstein-Glaser causal method in order to have a mathematical rigorous description of the free propagators of the theory, allowing us
to discuss a general treatment for propagators of the type $\left( k^{+}\right) ^{-n}$. Afterwards,
we show that at given conditions our results reproduce known prescriptions in the literature.
\end{abstract}

\newpage

In 1949 Dirac \cite{1} showed that different choices of the time evolution parameter \footnote{Given by the light-front $\{x^{\pm} \sim x^{0}\pm x^{3}\}$ or the usual instant-form $\{x^{0}\}$, and they are not related by a Lorentz transformation of coordinates.} are possible and that this can drastically change the content and interpretation of a given theory. However, a dynamical physical theory when written in the light-front form it becomes severely constrained with many
second-class constraints. These can be eliminated by constructing the generalized Dirac brackets, making it possible to develop a canonical quantization by the correspondence principle in terms of a reduced number of independent fields \cite{34}. Moreover, the light-front quantization \cite{35} is in fact very economical in displaying the relevant degrees of freedom, the discussion of the physical Hilbert space and the vacuum becomes more tractable. One may even say that the main advantage of the light-front quantization is the apparent simplicity of the vacuum state \cite{30}, where the physical vacuum is trivial. Many other interesting features were noticed by several authors, for instance, in the analysis of nonperturbative effects in the context of QCD \cite{37}, which prompted gradually the interest in the study of the
light-front form field theory as proposed by Dirac.

Now regarding the study of gauge field theories in the light-front formulation we may cite the original attempts at setting up the canonical quantization of QED in the light-cone gauge $A_{-}=0$, which has been known for almost forty years \cite{31,40,41}. The light-cone gauge was also used to quantize the Yang-Mills theory and in the analysis of its canonical structure and Dirac brackets, since it simplifies greatly the treatment of the constraints of gauge fields \cite{9}. We may also cite as a more intriguing application of this framework the evaluation of quantum effects contributing to the leading logarithm approximation in deep-inelastic processes \cite{32,33}. However, difficulties and inconsistencies remain in the quantization, some problems were associated with such gauge choice: Feynman amplitudes at the one-loop level exhibited double-pole singularities \cite{33}. This pathological behavior has been ascribed to the Principal Value (PV) prescription employed to the treatment the poles $\left( k.n\right) ^{-1}$ of the gauge boson propagator \cite{2,3}.

Later on, Mandelstam \cite{4} and Leibbrandt \cite{5} independently authored two prescriptions that circumvented the pathology above: $\left( k.n\right) ^{-1}$ singularities in the light-cone gauge. This prescription, Mandelstam-Leibbrandt, has been exhaustively and successfully tested, allowing for a suitable form to handle those singular factor ensuing in the light-cone gauge. However, it should be emphasized that the definition of higher powers of the singularity is not settled by this prescription. Nevertheless, over a decade later, Pimentel and Suzuki in \cite{7} revisited the PV prescription and assigned to the aforementioned failure of the PV program to the fact that when it is naively employed it violates causality. This motivated the proposal of a new and rather natural prescription, starting from the premise that the propagator as a whole must be \textit{causal} to treat the light-cone pole (also the higher-order poles), they proposed a new prescription known as \textit{Pimentel-Suzuki} prescription. Showing, therefore, that mathematics only does not suffice for such a task. A clear advantage of the causal prescription is the simplicity in performing the relevant integrals \cite{8}.

Here we would like to have a fresh look at the aforementioned light-cone poles pathology. It is not correct to say that it would be revisited by proposing a new prescription; but rather revisiting the pathology by analysing it in a natural and general framework where one can handle the singularities properly, i.e., where a prescription is not necessary. Our development will consist in a distributional approach, more specifically, we will make use of the strength of the analytic representation of distributions and axiomatic approach \cite{10,11,12,13,14}. In fact it was emphasized earlier in \cite{6} that in order to give a meaning to powers of these singularities the distributional nature of the Green's functions has to be taken into account. Hence, we will show how the analytic representation arises naturally by demanding that general properties, such as causality, be mandatory. The program will consist in two parts: in this first analysis we will show how to define the positive and negative frequency propagator to the scalar, fermionic and gauge fields in the light-cone form in a systematic and natural way. The approach that we show fit into the Wightman's formalism \cite{10}. To fix some ideas we review the case of the massive scalar field in the Wightman framework. We only require the minimum necessary for this theory to give the Cauchy integral representation. Then we show how to extend the results to the light-front form. Consequently, it will be constructed the equal-time (anti)commutators of the dynamical relevant fields and show that they reproduce known results. Though this first part of the development may look like as an exercise, but it has as motivation to show and present how powerful and simple the analytic representation for propagators may be, and how it points towards the use of the distributional machinery \cite{15,16} to deal with more intriguing quantities, this leads to the introduction of the general features from the \emph{causal} method proposed by Epstein and Glaser \cite{18}. This method was formulated in order to give a mathematical rigorous treatment of ultraviolet divergences in quantum field theory. In such framework such divergences do not appear anywhere in the calculations due to the correct splitting of the causal distributions into its advanced and retarded parts \cite{18,19,20}. Due to the properties of finiteness of the causal approach we expect to present an answer regarding the issue observed in \cite{36} that no complete regularization of the singularity is achieved and the presence of non-local ultraviolet terms show up in loop diagrams when the Mandelstam-Leibbrandt prescription is used. The fruits of the whole analysis will be concentrated mainly in dealing, through the Epstein-Glaser's causal method, for a general description of poles of the form
\begin{equation}
g\left(k;n\right)=\frac{1}{\left( k^{+}\right) ^{n}}  \label{eq 0.0}
\end{equation}%
where $n\geq1$. We will show that it is not needed to rely on prescriptions to deal to that, but generally with operator-valued distributions \cite{19,20}. Actually, there are further interesting studies in this direction \cite{22,23,24}. Nevertheless, some words may be spend about the poles \eqref{eq 0.0}. These poles have some problems such as the Wick rotation \cite{17} is not allowed. Moreover, the aforementioned prescriptions were designed in order to ensure that the location of the poles in the $k^{0}$-plane -- located in the second and fourth quadrants -- would not hinder Wick rotation nor spoil power-counting \cite{4,5,7}. It is worth to remark that the Wick rotation is a technique to define a particular distribution: the Feynman propagator. This method consists in defining distributions as boundary values of analytic functions \cite{8}. We believe that the Wick rotation technique fails in dealing to poles of the form \eqref{eq 0.0} because it is grounded into the distributional form only but not in the general principles, such as causality. The investigation is completed in a second paper \cite{49}, which contains a rather detailed exposition of the causal approach and discuss the radiative correction for the light-front QED, in particular, the vacuum polarization tensor.

In this paper, we revisit the light-cone poles pathology by studying the analytic representation of the positive and negative frequency propagator in order to derive the equal-time (anti)commutation relations, and subsequently by making use of the Epstein-Glaser's causal approach to construct the Feynman propagator; in particular, we discuss the general expression, in the causal framework, to dealing with the light-cone poles $1/(k^{+})^{n}$. We start by reviewing the general properties of the Wightman's formalism and showing constructively how the analytic representation emerges when one claims a physical principle, such as spectral condition, for a scalar field in the Sect.\ref{sec:1}. Next, in Sect.\ref{sec:2}, we make use of the analytic representation to derive consistently the general positive and negative frequency propagators for the scalar, fermionic and gauge fields in the light-front. Consequently, in possessing of these results, we derive the equal-time (anti)commutation relations to the dynamical fields in Sect.\ref{sec:3}. In Sect.\ref{sec:4}, we review the main aspects of the causal approach, and discuss in details, by considering a scalar field, the major role played by the splitting solution for regular and singular distributions in defining the retarded distribution. Moreover, we determine the expression for the Feynman propagator for the fermionic and gauge field; in particular, we find, without using a prescription, the expression of the photon propagator in the light-front, showing explicitly how it occurs naturally the presence of the proper contour for the light-front poles in the $k^{+}$-plane. In Sect.\ref{sec:5}, we make use of the results obtained in the previous sections in order to discuss the general expression of the propagator associated to the poles $1/(k^{+})^{n}$ in the framework of the causal approach. In Sect.\ref{sec:6} we summarize the results, and present our final remarks and prospects.


\section{Analytic representation for propagators}

\label{sec:1}

As we have mentioned earlier the Wightman's formalism is an axiomatic field theory \cite{10}. This is given at the beginning by stating general principles in the form of postulates, this approach guarantees that these principles are always obeyed. One may also refer to them as the Wightman's axioms.

Before starting with our developments, we will briefly review some points of the formalism that are important to our purpose. We start, by simplicity, from the general solution of the Klein-Gordon equation $\left( \square +m^{2}\right) \phi \left( x\right) =0$. The theory is formulated in terms of a set of covariant operator-valued distributions, $\phi $, which generates the full Hilbert space from the invariant vacuum $\left\vert \Omega \right\rangle $. We remark that for free fields we have a general distributional solution,%
\begin{equation}
\phi \left( x\right) =\left( 2\pi \right) ^{-2}\int d^{4}k\delta \left(
k^{2}-m^{2}\right) \tilde{a}\left( k\right) e^{-ikx},  \label{eq 0.3}
\end{equation}%
since we have considered $\phi $ hermitian, then $\tilde{a}\left( -k\right) =\tilde{a}^{\dag }\left( k\right) $. In this formalism the fields $\phi \left(x\right) $ are not functions but operator-valued distributions, this means that for each test function $f\left( x\right) $ it is associated an operator $\left\langle \phi ,f\right\rangle $. Usually in the Wightman's formalism it is adopted the notation $\phi \left[ f\right] =\left\langle \phi ,f\right\rangle $, and defined as \footnote{This is only possible for regular distributions, but not for singular ones.}%
\begin{equation}
\phi \left[ f\right] =\int d^{4}x\phi \left( x\right) f\left( x\right) .
\label{eq 0.4}
\end{equation}%
To guarantee the existence of the Fourier transformation of distributions, it is considered the Schwartz space $\mathbf{S}\left( \mathbb{R}^{4\mathbf{n}}\right) $ of the test functions $f$ \cite{10}. In this space we can define the
Fourier transformation of the scalar field, $\hat{\phi}\left( k\right) $, as it follows%
\begin{equation}
\left\langle \phi\left(x\right),f \left(x\right)\right\rangle = \left\langle \hat{\phi}\left(k\right),\check{f}\left(k\right)\right\rangle =\int dk\hat{\phi}\left( k\right) \check{f}\left( k\right) ,\label{eq 0.5}
\end{equation}%
where $\check{f}$ is the inverse Fourier transformation of $f$, Eq.~\eqref{A.2}, and is a well-behaved test function of $\hat{\phi}$. The first equality of \eqref{eq 0.5} followed by the Parseval theorem. Besides, $\phi$ is contained in the dual space $\mathbf{S}^{\prime }\left( \mathbb{R}^{\mathbf{m}}\right)$, and for this reason we say that $\phi$ is an operator-valued \textit{tempered distribution}. Furthermore, it is convenient to split $\phi$ into its positive and negative frequency components%
\begin{equation}
\phi \left( x\right) =\phi ^{\left( +\right) }\left( x\right) +\phi ^{\left(
-\right) }\left( x\right) ,
\end{equation}%
where $\phi ^{\left( +\right) }$ is named the positive and $\phi ^{\left(-\right) }$ the negative part of the field. We obtain that their explicit expressions, after some algebraic manipulation, are defined by the relations%
\begin{align}
\phi ^{\left( -\right) }\left[ f\right] \left\vert \Omega \right\rangle
&= \int d^{4}k\theta \left( k_{0}\right) \delta \left( k^{2}-m^{2}\right)
\tilde{a}\left( k\right) \hat{f}\left( -k\right) \left\vert \Omega
\right\rangle ,  \label{eq 0.8} \\
\phi ^{\left( +\right) }\left[ f\right] \left\vert \Omega \right\rangle
&= \int d^{4}k\theta \left( k_{0}\right) \delta \left( k^{2}-m^{2}\right)
\tilde{a}^{\dag }\left( k\right) \hat{f}\left( k\right) \left\vert \Omega
\right\rangle .  \label{eq 0.9}
\end{align}%
By the \textit{spectral condition} \footnote{No states of negative-energy exists, i.e., the eigenvalues of the operator $P_{\mu}$ $\left( P^2=m^2\right)$ lie in on the plus cone $V^{+}\left( k\right)$ \cite{10}.}, we do not have components satisfying $k \in \bar{V}^{-}\left( k\right)\rightarrow -k\in \bar{V}^{+}\left( k\right)$ (see \eqref{eq 4.2a}),
then the part associated to $\hat{f}\left( -k\right) $ must be zero, which is satisfied if%
\begin{equation}
\tilde{a}\left( k\right) \left\vert \Omega \right\rangle =0,
\end{equation}%
whereas, since the nonzero components are in $k \in \bar{V}^{+}\left( k\right)$, the part associated to $\hat{f}\left( k\right) $ must be nonzero, which is satisfied if%
\begin{equation}
\tilde{a}^{\dag }\left( k\right) \left\vert \Omega \right\rangle \neq 0.
\end{equation}%
From these conditions follow that the operators $\tilde{a}\left( k\right) $ and $\tilde{a}^{\dag }\left( k\right) $ are called as the operators of annihilation and creation, respectively.

In this formalism the central objects are the so-called Wightman's functions. They are defined as the vacuum expectation values (vev) of a product of fields. For instance, the $n$-points Wightman's function for scalar fields is given by%
\begin{equation}
W_{n}\left( x_{1},x_{2},\ldots ,x_{n}\right) =\left\langle \Omega
\right\vert \phi \left( x_{1}\right) \phi \left( x_{2}\right) \ldots \phi
\left( x_{n}\right) \left\vert \Omega \right\rangle .  \label{eq 1.0}
\end{equation}%
Of course they are not functions in the strict sense, but tempered distributions \cite{10},
\begin{equation}
W_{2}\left( x_{1}-x_{2}\right) =\left( 2\pi \right) ^{-2}\int d^{4}k\hat{W}%
_{2}\left( k\right) e^{-ik\left( x_{1}-x_{2}\right) },  \label{eq 1.1}
\end{equation}%
where $\hat{W}_{2}\left( k\right) $ is the two-point Wightman's function in the momentum space. Moreover, we have that for free scalar fields, the Wightman's function obeys the same equation of the free scalar field,
\begin{equation}
\left( \square _{i}+m^{2}\right) W_{2}\left( x_{1}-x_{2}\right) =0,\quad i=1,2.
\end{equation}%
Hence, as a consequence of the spectral condition, one can find that the two-point Wightman function in the momentum space is given by
\begin{equation}
\hat{W}_{2}\left( k\right) =\frac{1}{2\pi }\theta \left( k_{0}\right) \delta
\left( k^{2}-m^{2}\right) .  \label{eq 1.2}
\end{equation}%
Hence, with the physical concepts and necessary tools in hands, we shall now introduce the analytic representation of propagators. In order to elucidate the content we shall discuss the case of scalar fields first, to only then introduce the spinor and vector fields.

\subsection{Analytic representation of the scalar propagator}

Once the fundamental propagators are linear combinations of the positive (PF) and negative (NF) frequency parts of the propagator, it is rather natural to consider them here in our development. We define the PF propagator by the relation from the contraction between scalar fields:%
\begin{equation}
\overbrace{\phi \left( x\right) \phi \left( y\right) }\equiv \left[ \phi
^{\left( -\right) }\left( x\right) ,\phi ^{\left( +\right) }\left( y\right) %
\right] =-iD_{m}^{\left( +\right) }\left( x-y\right) .  \label{eq 1.10}
\end{equation}%
Moreover, for a normalized vacuum, we have that this propagator can also be written as it follows%
\begin{equation}
D_{m}^{\left( \pm \right) }\left( x-y\right) =i\left\langle \Omega \left\vert
\left[ \phi ^{\left( \mp \right) }\left( x\right) ,\phi ^{\left( \pm \right)
}\left( y\right) \right] \right\vert \Omega \right\rangle .  \label{eq 1.11}
\end{equation}%
Now, by using the properties of the positive and negative parts of the field, Eqs.\eqref{eq 0.8} and \eqref{eq 0.9}, we have that%
\begin{equation}
D_{m}^{\left( +\right) }\left( x-y\right) =i\left\langle \Omega \left\vert
\phi \left( x\right) \phi \left( y\right) \right\vert \Omega \right\rangle .
\label{eq 1.12}
\end{equation}%
Finally, it is not difficult to obtain the relation between the PF and NF propagators and the two-point Wightman's function \eqref{eq 1.0}%
\begin{equation}
D_{m}^{\left( -\right) }\left( x-y\right) =-D_{m}^{\left( +\right) }\left(
y-x\right) =-iW_{2}\left( y-x\right)  \label{eq 1.14a}
\end{equation}%
Hence, with the above results we can make use of the expression \eqref{eq 1.2} to thus obtain the PF and NF propagators written in the momentum space%
\begin{equation}
\hat{D}_{m}^{\left( \pm \right) }\left( k\right) =\pm \frac{i}{2\pi }\theta
\left( \pm k_{0}\right) \delta \left( k^{2}-m^{2}\right) ,  \label{eq 1.15}
\end{equation}%
moreover, we can understand \eqref{eq 1.15} as distributions in $k^{2}$, but if we write them equivalently as%
\begin{equation}
\hat{D}_{m}^{\left( \pm \right) }\left( k\right) =\pm \frac{i}{2\pi }\theta
\left( \pm k_{0}\right) \delta \left( k_{0}^{2}-\omega _{m}^{2}\right) =%
\frac{i}{2\pi }\frac{\delta \left( k_{0}\mp \omega _{m}\right) }{k_{0}\pm
\omega _{m}},  \label{eq 1.16}
\end{equation}%
where $\omega _{m}=\sqrt{\vec{k}^{2}+m^{2}}$ is the frequency, then these \eqref{eq 1.16} can be understood either as distributions in $k_{0}$. Therefore, with the previous results we are able to find the analytic representation of the propagator. For this purpose we can make use of the following definition for the $\delta $-Dirac translated distribution 
\begin{equation}
\varphi \left( \pm \omega _{m}\right) =\left\langle \delta \left( k_{0}\mp
\omega _{m}\right) ,\varphi \left( k_{0}\right) \right\rangle ,
\label{eq 1.17}
\end{equation}%
where $\varphi $ is a test function. Hence, the propagators $\hat{D}_{m}^{\left( \pm \right) }$ can be defined by the following functional relation
\begin{equation}
\left\langle \hat{D}_{m}^{\left( \pm \right) },\varphi \right\rangle =\frac{i%
}{2\pi }\left[ \frac{\left\langle \delta \left( k_{0}\mp \omega _{m}\right)
,\varphi \left( k_{0}\right) \right\rangle }{k_{0}\pm \omega _{m}}\right]
=\left( 2\pi \right) ^{-2}\left\{ 2\pi i\left[ \frac{\varphi \left(
k_{0}\right) }{k_{0}\pm \omega _{m}}\right] _{k_{0}=\pm \omega _{m}}\right\}.
\end{equation}%
Besides, identifying the Cauchy integral in the $k_{0}$-complex plane, we are finally able to obtain the analytic representation of the PF and NF scalar propagators as being%
\begin{equation}
\left\langle \hat{D}_{m}^{\left( \pm \right) },\varphi \right\rangle =\left(
2\pi \right) ^{-2}\oint\limits_{c_{\pm }}\frac{\varphi \left( k_{0}\right) }{%
k_{0}^{2}-\omega _{m}^{2}}dk_{0},  \label{eq 1.18}
\end{equation}%
where $c_{+\left( -\right) }$ is a counterclockwise closed path which contains only the positive (negative) poles of the Green's function $\hat{g}\left( k\right) =\frac{1}{k_{0}^{2}-\omega _{m}^{2}}$.

Before concluding this section, it is interesting as to our next development to present useful remarks here. With the PF and NF propagators we may find the propagator:%
\begin{equation}
D_{m}\left( x\right) =D_{m}^{\left( + \right) }\left( x\right)+D_{m}^{\left( - \right) }\left( x\right),  \label{eq 4.2}
\end{equation}%
named \emph{causal} propagator because it has causal support, i.e. it vanishes outside the closed forward and backward light-cone
\begin{equation}
\text{Supp}~ D_{m}\left( x\right) \subseteq \bar{V}^{-}\left( x\right)+\bar{V}^{+}\left( x\right), \quad \bar{V}^{\pm }\left( x\right)=\left\{ x | \text{ }x^{2}\geq 0,\quad \pm x_{0}\geq
0\right\} .  \label{eq 4.2a}
\end{equation}%
Moreover, the causal propagator \eqref{eq 4.2} can be split into two important propagators: one which indicates the propagation to the future and another to the past. These are the so-called \emph{retarded} and \emph{advanced} propagator which vanishes for $x_{0}<0$ and $x_{0}>0$, respectively, in whatever referential. They are related to the causal propagator as follows%
\begin{equation}
D_{m}^{R}\left( x\right) =\theta \left( x_{0}\right) D_{m}\left( x\right) ,
\quad D_{m}^{A}\left( x\right) =-\theta \left( -x_{0}\right) D_{m}\left( x\right) .
\label{eq 4.5}
\end{equation}%
Another important distributional solution is the so-called \emph{Feynman} propagator%
\begin{equation}
D_{m}^{F}\left( x\right) =\theta \left( x_{0}\right) D_{m}^{\left( +\right)
}\left( x\right) -\theta \left( -x_{0}\right) D_{m}^{\left( -\right) }\left(
x\right) ,  \label{eq 4.6}
\end{equation}%
which is related to the vacuum expectation value of time-ordered product of fields. Moreover, this distribution can be written as the Fourier transformation%
\begin{equation}
\hat{D}_{m}^{F}\left( k\right) =-\left( 2\pi \right) ^{-2}\frac{1}{\left[
\left( k_{0}-i0^{+}\right) -\left( -\omega _{m}\right) \right] \left[ \left(
k_{0}+i0^{+}\right) -\left( \omega _{m}\right) \right] }.  \label{eq 4.7}
\end{equation}%
Also, it can be understood as the boundary value of the following complex analytic function%
\begin{equation}
\hat{D}_{m}^{F}\left( k\right) =-\left( 2\pi \right) ^{-2}\lim_{\eta
\rightarrow 0^{+}}\frac{1}{\left[ \left( k_{0}-i\eta \right) -\left( -\omega
_{m}\right) \right] \left[ \left( k_{0}+i\eta \right) -\left( \omega
_{m}\right) \right] },  \label{eq 4.8}
\end{equation}%
this is the definition of the so-called Wick rotation technique.

Another equivalent way to write the Feynman propagator is by using the definition of the retarded or advanced distribution \eqref{eq 4.5},%
\begin{equation}
D_{m}^{F}\left( x\right) =D_{m}^{R}\left( x\right) -D_{m}^{\left( -\right)
}\left( x\right) =D_{m}^{A}\left( x\right) +D_{m}^{\left( +\right) }\left(
x\right) .  \label{eq 4.9}
\end{equation}%
This is not a superfluous equivalence to the Wick rotation. On the other hand, when we separate it in the positive and negative part one finds%
\begin{align}
D_{m}^{F\left( +\right) }\left( x\right) = D_{m}^{R\left( +\right) }\left(
x\right) ,  \quad D_{m}^{F\left( -\right) }\left( x\right) = D_{m}^{A\left( -\right) }\left(
x\right) ,  \label{eq 4.11}
\end{align}%
and also use the definition of the retarded and advanced propagators, we can show that the Feynman propagator has the following causal property: \emph{Only positive-frequency solution can be propagating to the future and only negative-frequency solution can be propagating to the past} \cite{21}. Thus the relation \eqref{eq 4.9} and the general definition of the different propagators are the starting point of our axiomatic approach.

Therefore in possessing of the basic results regarding the analytic representation of a propagator, we are now ready for the subsequent development. We shall proceed in evaluating the basic commutators for the dynamical fields in the light-front, but first we shall derive the respective propagators for the scalar, spinor and gauge fields.


\section{Light-front propagators}

\label{sec:2}

The PF and NF propagators are distributional solutions of the free field equations, then any linear combination of these is also a solution; for example, we may define the \emph{causal} propagator distributional solution: $D=D^{\left( +\right) }+D^{\left( -\right) }$ as in \eqref{eq 4.2}. Equivalently, we can write it in the momentum space%
\begin{equation}
\hat{D}\left( k\right) =\hat{D}^{\left( +\right) }\left( k\right) +\hat{D}%
^{\left( -\right) }\left( k\right) ,  \label{eq 1.22}
\end{equation}%
and from \eqref{eq 1.15} it follows that its support in the momentum space is contained in:%
\begin{equation}
\text{Supp}~\hat{D}\left( k\right) =\text{Supp}~\hat{D}^{\left( +\right) }\left( k\right)
\cup \text{Supp}~ \hat{D}^{\left( -\right) }\left( k\right) =\bar{V}^{+}\left(
k\right) \cup \bar{V}^{-}\left( k\right) .  \label{eq 1.23}
\end{equation}%
In order to implement our analysis of the light-front dynamics, it is interesting to generalize the previous result \eqref{eq 1.18} for any dynamics form \cite{1} as it will become clear next. We notice that from the analytic representation of the PF and NF propagators Eq.~\eqref{eq 1.18} and expression \eqref{eq 1.22} we obtain the scalar propagator
\begin{equation}
\left\langle \hat{D}_{m},\varphi \right\rangle =\left(
2\pi \right) ^{-2}\left(\oint\limits_{c_{+}}+\oint\limits_{c_{-}}\right)\frac{\varphi \left( k_{0}\right) }{%
k_{0}^{2}-\omega _{m}^{2}}dk_{0}, 
\end{equation}%
Now, we can generalize this result for an arbitrary propagator, such as follows%
\begin{equation}
\left\langle \hat{D},\varphi \right\rangle =\left( 2\pi \right)
^{-2}\oint\limits_{c_{all}}\hat{G}\left( k\right) \varphi \left(
k_{0}\right) dk_{0},  \label{eq 1.19}
\end{equation}%
where $c_{all}$ are all counterclockwise closed paths which contain all individual poles and $\hat{G}\left( k\right) $ is the Green's function associated to the free field equation. However, to return to the PF and NF propagators from this quantity, it is only necessary to split correctly its support into the closed forward $\bar{V}^{+}$ and closed backward $\bar{V}^{-}$ cone, respectively. Hence, in order to implement this idea we may introduce into the analytic representation \eqref{eq 1.19} a time-like or light-like curve such that it crosses the origin. If we define $k_{\lambda }$ as the parameter of this curve, such that $k_{\lambda }=0$ corresponds to the origin, the splitting of the supports can be made with this single parameter as follows%
\begin{equation}
\left\langle \hat{D}^{\left( \pm \right) },\varphi \right\rangle =\left(
2\pi \right) ^{-2}\theta \left( \pm k_{\lambda }\right)
\oint\limits_{c_{all}}\hat{G}\left( k\right) \varphi \left( k_{0}\right)
dk_{0}.  \label{eq 1.20}
\end{equation}%
Let us discuss some further properties of this last expression. First, notice that $k_{0}$ is the variable in which the poles of the Green's function are expressed, thus the poles can be interpreted as the cuts over the support of $ \hat{D} $ (which are surfaces in the Minkowski space) by a time-like curve, parametrized by $k_{0}$. Nevertheless, since the support of $ \hat{D} $ is embedded into $ \bar{V}^{+}\left(k\right) \cup \bar{V}^{-}\left( k\right) $, Eq.~\eqref{eq 1.22}, then we can choose any other arbitrary time-like curve or even a light-like curve parametrized by $k_{\sigma }$. Therefore the expression \eqref{eq 1.20} takes the following general form%
\begin{equation}
\left\langle \hat{D}^{\left( \pm \right) },\varphi \right\rangle =\left(
2\pi \right) ^{-2}\theta \left( \pm k_{\lambda }\right)
\oint\limits_{c_{all}}\hat{G}\left( k\right) \varphi \left( k_{\sigma
}\right) dk_{\sigma }.  \label{eq 1.21}
\end{equation}%
At first sight $k_{\lambda }$ seems to be arbitrary, but we must avoid those choices that are ill-defined in the distributional sense, e.g., $k_{\lambda }=k_{\sigma }$, which may lead to a ill-defined distributional product, for instance $\theta \left( k_{\lambda }\right) \delta \left(k_{\lambda }\right) $ \cite{bp}.

It should be emphasized that the parameter $ k_{\sigma} $ only plays the role in pointing out the Green's functions poles, whereas the parameter $k_{\lambda }$ takes the role to split the propagator in its positive and negative frequency part, thus, $k_{\lambda }$ is related to the energy variable. In particular, in light-front dynamics it is usually taken the temporal variable to be $ x^{+} $, so the energy must be indicated by $ k^{-} $, and in order to avoid any ill-defined distributional product, we may choose as the pole parameter to be $ k^{+} $. Thus, in light-front dynamics the expression \eqref{eq 1.21}  takes the following form
\begin{equation}
\left\langle \hat{D}^{\left( \pm \right) },\varphi \right\rangle =\left(
2\pi \right) ^{-2}\theta \left( \pm k^{-}\right)
\oint\limits_{c_{all}}\hat{G}\left( k\right) \varphi \left( k^{+}\right) dk^{+}.  \label{eq 1.23a}
\end{equation}%

\subsection{Scalar propagators}

Let us start the discussion by the massive scalar field $\phi \left(x\right) $. It has already been discussed that the scalar free field satisfies the equation of motion: $\left( \square +m^{2}\right) \phi \left( x\right) =0$. Then, its corresponding Green's function is given by%
\begin{equation}
\hat{G}_{m}\left( p\right) =\frac{1}{p^{2}-m^{2}}.
\end{equation}%
Moreover, we can rewrite this Green's function explicitly in terms of the light-front coordinates \footnote{Our notation to the light-front coordinates is presented in the \ref{sec:appA}.}%
\begin{equation}
\hat{G}_{m}\left( p\right) =\frac{1}{2p^{+}p^{-}-\omega _{m}^{2}},
\label{eq 2.1}
\end{equation}%
where the frequency $\omega _{m}$ is now written as: $\omega _{m}=\sqrt{p_{\bot }^{2}+m^{2}}$. Next we should choose a convenient coordinate to express the poles, in this case we can choose either $p^{+}$ or $p^{-}$. As we have explained above, it is convenient for our purposes to choose $p^{+}$ as distributional variable. From that it follows that the analytic representation of the PF and NF propagator \eqref{eq 1.23a} can be written as%
\begin{equation}
\left\langle \hat{D}_{m}^{\left( \pm \right) },\varphi \right\rangle =\left(
2\pi \right) ^{-2}\theta \left( \pm p^{-}\right) \frac{1}{2p^{-}}%
\oint\limits_{c_{all}}\frac{\varphi \left( p^{+}\right) }{\left( p^{+}-\frac{%
\omega _{m}^{2}}{2p^{-}}\right) }dp^{+},  \label{eq 2.2}
\end{equation}%
in which we have chosen as the split parameter the variable $p^{-}$, and thus $c_{all}$ are all counterclockwise closed paths which contain all individual poles in the complex plane of $p^{+}$. Moreover, by means of some distributional properties we obtain%
\begin{equation}
\left\langle \hat{D}_{m}^{\left( \pm \right) },\varphi \right\rangle =\left(
2\pi \right) ^{-2}\theta \left( \pm p^{-}\right) \frac{2\pi
i}{2p^{-}}\left\langle \delta \left( p^{+}-\frac{\omega _{m}^{2}}{2p^{-}}\right)
,\varphi \left( p^{+}\right) \right\rangle .
\end{equation}%
By comparing both sides one gets
\begin{equation}
\hat{D}_{m}^{\left( \pm \right) }\left( p\right) =\theta \left( \pm
p^{-}\right) \frac{i}{4\pi p^{-}}\delta \left( p^{+}-\frac{\omega
_{m}^{2}}{2p^{-}}\right) .  \label{eq 2.3}
\end{equation}%
Finally, we obtain the light-front scalar PF and NF propagators%
\begin{equation}
\hat{D}_{m}^{\left( \pm \right) }\left( p\right) =\pm \frac{i}{2\pi }\theta
\left( \pm p^{-}\right) \delta \left( 2p^{+}p^{-}-\omega _{m}^{2}\right)
=\pm \frac{i}{2\pi }\theta \left( \pm p^{-}\right) \delta \left(
p^{2}-m^{2}\right) .  \label{eq 2.4}
\end{equation}

\subsection{Fermionic propagators}

The discussion for the fermionic fields follows by the same lines as for the scalar fields. Thus, the free Dirac spinors $\psi $ and $\bar{\psi}$ satisfy the free Dirac equations%
\begin{equation}
\left( i\gamma .\partial -m\right) \psi =0,\text{\quad }\bar{\psi}\left(
i\gamma .\overleftarrow{\partial }+m\right) =0.  \label{eq 2.5}
\end{equation}%
Without any complication one obtains the fermionic Green's function%
\begin{equation}
\hat{S}\left( p\right) =\left( \gamma .p+m\right) \hat{G}_{m}\left( p\right)
,  \label{eq 2.5a}
\end{equation}%
where $\hat{G}_{m}\left( p\right) $ is the scalar Green's function \eqref{eq 2.1}. Now, if we choose $p^{+}$ as the pole parameter then we can write $\hat{S}\left( p\right) $ as
\begin{equation}
\hat{S}\left( p\right) =\frac{\left( \gamma .p+m\right) }{\left(
2p^{-}\right) \left( p^{+}-\frac{\omega _{m}^{2}}{2p^{-}}\right) }.
\label{eq 2.6}
\end{equation}%
We see clearly that the factor $\left( \gamma .p+m\right) $ does not cancel any poles, then the fermionic PF and NF propagators are given by%
\begin{equation}
\hat{S}^{\left( \pm \right) }\left( p\right) =\left( \gamma .p+m\right) \hat{%
D}_{m}^{\left( \pm \right) }\left( p\right),
\end{equation}%
where $\hat{D}_{m}^{\left( \pm \right) }\left( p\right) $ are the scalar PF and NF propagator given in the previous section, Eq.~\eqref{eq 2.4}.

\subsection{Electromagnetic propagators}

In the analysis of the free gauge field in the light-front, we will make use of a previous result as found in \cite{9}, where it was shown explicitly the presence of two Lagrange multipliers $\left( \partial .A\right) ^{2}$ and $\left( \eta .A\right) ^{2}$ in the usual free electromagnetic Lagrangian density. It follows then the complete expression%
\begin{equation}
\mathcal{L}=-\frac{1}{4}F_{\mu \nu }F^{\mu \nu }-\frac{1}{2\beta }\left(
\partial _{\mu }A^{\mu }\right) ^{2}-\frac{1}{2\alpha }\left( \eta _{\mu
}A^{\mu }\right) ^{2},  \label{eq 2.7}
\end{equation}%
where $\alpha $, $\beta $ are arbitrary constants. From this Lagrangian we obtain the following free field equation:%
\begin{equation}
\left[ \left( \square h_{\mu \nu }-\partial _{\mu }\partial _{\nu }\right) +%
\frac{1}{\beta }\partial _{\mu }\partial _{\nu }-\frac{1}{\alpha }\eta _{\mu
}\eta _{\nu }\right] A^{\nu }=0.  \label{eq 2.8}
\end{equation}%
Thus, we have that the free Green's function is given by%
\begin{align}
\hat{G}_{\mu \nu }\left( k\right) &= h_{\mu \nu }\frac{1}{k^{2}}+\frac{%
\left( \beta -1\right) \left( \alpha k^{2}+\eta ^{2}\right) }{\left[ \left(
\beta -1\right) \left( k.\eta \right) ^{2}+\left( \alpha k^{2}+\eta
^{2}\right) k^{2}\right] k^{2}}k_{\mu }k_{\nu }\notag\\
&\quad-\frac{\left( \beta -1\right) \left( k.\eta \right) }{\left[ \left( \beta
-1\right) \left( k.\eta \right) ^{2}+\left( \alpha k^{2}+\eta ^{2}\right)
k^{2}\right] k^{2}}\left( k_{\mu }\eta _{\nu }+k_{\nu }\eta _{\mu }\right) \notag\\
&\quad  -\frac{1}{\left( \beta -1\right) \left( k.\eta \right) ^{2}+\left( \alpha
k^{2}+\eta ^{2}\right) k^{2}}\eta _{\mu }\eta _{\nu }. \label{eq 2.8a}
\end{align}%
Nevertheless, it is rather interesting to consider, as a particular case, the propagator in the light-front $\eta ^{2}=0$, as well as in the two transverse conditions: $\hat{G}_{\mu \nu }k^{\mu }=0$ and $\hat{G}_{\mu \nu }\eta ^{\mu }=0$. Therefore, one may rewrite the Green's function \eqref{eq 2.8a} as the following%
\begin{equation}
\hat{G}_{\mu \nu }\left( k\right) =\frac{h_{\mu \nu }}{k^{2}}-\frac{k_{\mu
}\eta _{\nu }+k_{\nu }\eta _{\mu }}{\left( k.\eta \right) k^{2}}+\frac{\eta
_{\mu }\eta _{\nu }}{\left( k.\eta \right) ^{2}}.  \label{eq 2.9}
\end{equation}%
Finally, choosing in particular: $\eta ^{\mu }=\left( 0,0,0,1\right) $, thus: $k.\eta =k_{-}=k^{+}$; it then follows the expression for the free Green's function%
\begin{equation}
\hat{G}_{\mu \nu }\left( k\right) =\frac{h_{\mu \nu }}{k^{2}}-\frac{k_{\mu
}\eta _{\nu }+k_{\nu }\eta _{\mu }}{k^{2}k^{+}}+\frac{\eta _{\mu }\eta _{\nu
}}{\left( k^{+}\right) ^{2}}.  \label{eq 2.10}
\end{equation}%
In order to deal with the poles from Eq.~\eqref{eq 2.10} we may define the general expression and solve it explicitly%
\begin{equation}
\hat{G}_{0}\left( k;n,l\right) =\frac{1}{\left( k^{2}\right) ^{n}\left(
k^{+}\right) ^{l}},
\end{equation}%
for the cases: $\left( n,l\right) =\left( 1,0\right) $, $\left( 1,1\right) $, $\left( 0,2\right) $, we have also naturally chosen the pole parameter $k^{+}$.

\begin{enumerate}[label=\textbf{\roman*}]

\item For $\left( n,l\right) =\left( 1,0\right) $ we see that this is nothing more than the massless scalar Green's function \eqref{eq 2.4}, then taking $m=0$%
\begin{equation}
\hat{D}_{0}^{\left( \pm \right) }\left( k;1,0\right) =\hat{D}_{0}^{\left(
\pm \right) }\left( k\right) =\pm \frac{i}{2\pi }\theta \left( \pm
k^{-}\right) \delta \left( 2k^{+}k^{-}-\omega _{0}^{2}\right),
\label{eq 2.11}
\end{equation}%
where $\omega _{0}=\sqrt{k_{\bot }^{2}}$.

\item For $\left( n,l\right) =\left( 1,1\right) $ we have
\begin{equation}
\hat{G}_{0}^{\left( \pm \right) }\left( k;0,1\right) =\frac{1}{k^{2}k^{+}}.
\end{equation}

By making use of the analytic representation \eqref{eq 1.23a} one gets%
\begin{equation}
\left\langle \hat{D}_{0}^{\left( \pm \right) },\varphi \right\rangle =\left(
2\pi \right) ^{-2}\theta \left( \pm k^{-}\right) \frac{1}{2k^{-}}%
\oint\limits_{c_{all}}\frac{\varphi \left( k^{+}\right) }{\left( k^{+}-\frac{%
\omega _{0}^{2}}{2k^{-}}\right) \left( k^{+}\right) }dk^{+}.  \label{eq 2.12}
\end{equation}%
We can evaluate the Cauchy integral for each one of the poles
\begin{equation}
\left\langle \hat{D}_{0}^{\left( \pm \right) },\varphi \right\rangle =\frac{i%
}{2\pi }\theta \left( \pm k^{-}\right) \frac{1}{2k^{-}}\left\{ \left[ \frac{%
\varphi \left( k^{+}\right) }{k^{+}}\right] _{k^{+}=\frac{\omega _{0}^{2}}{%
2k^{-}}}+\left[ \frac{\varphi \left( k^{+}\right) }{k^{+}-\frac{\omega
_{0}^{2}}{2k^{-}}}\right] _{k^{+}=0}\right\} ,
\end{equation}%
and, after some distributional manipulation, we obtain,%
\begin{equation}
\left\langle \hat{D}_{0}^{\left( \pm \right) },\varphi \right\rangle =\frac{i%
}{2\pi }\theta \left( \pm k^{-}\right) \frac{1}{\omega _{0}^{2}}\left[
\left\langle \delta \left( k^{+}-\frac{\omega _{0}^{2}}{2k^{-}}\right)
,\varphi \left( k^{+}\right) \right\rangle -\left\langle \delta \left(
k^{+}\right) ,\varphi \left( k^{+}\right) \right\rangle \right] .
\end{equation}%
Finally, by comparing both sides, it follows%
\begin{equation}
\hat{D}_{0}^{\left( \pm \right) }\left( k;1,1\right) =\frac{i}{2\pi }\theta
\left( \pm k^{-}\right) \frac{1}{\omega _{0}^{2}}\left[ \delta \left( k^{+}-%
\frac{\omega _{0}^{2}}{2k^{-}}\right) -\delta \left( k^{+}\right) \right] ,
\label{eq 2.13}
\end{equation}%
or even%
\begin{equation}
\hat{D}_{0}^{\left( \pm \right) }\left( k;1,1\right) =\pm \frac{i}{2\pi }%
\theta \left( \pm k^{-}\right) \frac{2k^{-}}{\omega _{0}^{2}}\left[ \delta
\left( 2k^{+}k^{-}-\omega _{0}^{2}\right) -\delta \left( 2k^{+}k^{-}\right) %
\right] .  \label{eq 2.14}
\end{equation}

\item For $\left( n,l\right) =\left( 0,2\right) $ we have%
\begin{equation}
\hat{G}^{\left( \pm \right) }\left( k\right) =\frac{1}{\left( k^{+}\right)
^{2}}.
\end{equation}%
Then making use of the analytic representation%
\begin{equation}
\left\langle \hat{D}_{0}^{\left( \pm \right) },\varphi \right\rangle =\left(
2\pi \right) ^{-2}\theta \left( \pm k^{-}\right) \oint\limits_{c_{all}}\frac{%
\varphi \left( k^{+}\right) }{\left( k^{+}\right) ^{2}}dk^{+}.
\label{eq 2.15}
\end{equation}%
After performing the Cauchy integral for second order pole%
\begin{equation}
\left\langle \hat{D}_{0}^{\left( \pm \right) },\varphi \right\rangle =\left(
2\pi \right) ^{-2}\theta \left( \pm k^{-}\right) \left\{ \frac{2\pi i}{%
\left( 2-1\right) !}\varphi ^{\left( 1\right) }\left( 0\right) \right\} ,
\end{equation}%
which can also be rewritten as the following%
\begin{equation}
\left\langle \hat{D}_{0}^{\left( \pm \right) },\varphi \right\rangle =-\frac{%
i}{2\pi }\theta \left( \pm k^{-}\right) \left\langle \delta ^{\left(
1\right) }\left( k^{+}\right) ,\varphi \left( k^{+}\right) \right\rangle .
\end{equation}%
Finally, by comparing both sides, one finds%
\begin{equation}
\hat{D}_{0}^{\left( \pm \right) }\left( k;0,2\right) =-\frac{i}{2\pi }\theta
\left( \pm k^{-}\right) \delta ^{\left( 1\right) }\left( k^{+}\right) .
\label{eq 2.16}
\end{equation}
\end{enumerate}

Therefore, substituting the results Eqs.\eqref{eq 2.11}, \eqref{eq 2.14} and \eqref{eq 2.16} into the complete expression \eqref{eq 2.10}, one finds that the electromagnetic PF and NF propagators are written as%
\begin{equation}
\hat{D}_{\mu \nu }^{\left( \pm \right) }\left( k\right) =h_{\mu \nu }\hat{D}%
_{0}^{\left( \pm \right) }\left( k;1,0\right) -\left( k_{\mu }\eta _{\nu
}+k_{\nu }\eta _{\mu }\right) \hat{D}_{0}^{\left( \pm \right) }\left(
k;1,1\right) +\eta _{\mu }\eta _{\nu }\hat{D}_{0}^{\left( \pm \right)
}\left( k;0,2\right) ,  \label{eq 2.17}
\end{equation}%
or even in its explicit form%
\begin{align}
\hat{D}_{\mu \nu }^{\left( \pm \right) }\left( k\right) =& \pm \frac{i}{2\pi
}\theta \left( \pm k^{-}\right) \bigg\{ \delta \left( 2k^{+}k^{-}-\omega
_{0}^{2}\right) h_{\mu \nu } \notag \\
& \quad-\frac{2k^{-}}{\omega _{0}^{2}}\left[ \delta
\left( 2k^{+}k^{-}-\omega _{0}^{2}\right) -\delta \left( 2k^{+}k^{-}\right) %
\right] \left( k_{\mu }\eta _{\nu }+k_{\nu }\eta _{\mu }\right) \bigg\}\notag \\
&\quad-\frac{i}{2\pi }\theta \left( \pm k^{-}\right) \delta ^{\left( 1\right)
}\left( k^{+}\right) \eta _{\mu }\eta _{\nu } .  \label{eq 2.18}
\end{align}%
Now that we have determined all the PF and NF propagators for the scalar, fermionic and gauge fields, we are ready to proceed in evaluating the light-front commutators from the dynamical fields, showing how easily they are obtained when the theory's construction follows an axiomatic approach.


\section{Light-front commutators}

\label{sec:3}

To construct the equal-time (anti)commutation relations between the dynamical fields we should notice first that the PF and NF parts of the propagator, $D^{\left( \pm \right) }$, are related to the commutators between the positive and negative parts of the free field. Then, it is not difficult to show that the causal propagator, $D=D^{\left( +\right) }+D^{\left( -\right)} $, is related to the (anti)commutator between the free fields%
\begin{align}
\left[ \phi \left( x\right) ,\phi \left( y\right) \right] &= -iD_{m}\left(
x-y\right) ,  \label{eq 3.2} \\
\left\{ \psi \left( x\right) ,\bar{\psi}\left( y\right) \right\}
&= -iS\left( x-y\right) ,  \label{eq 3.3} \\
\left[ A_{\mu }\left( x\right) ,A_{\nu }\left( y\right) \right] &= iD_{\mu
\nu }\left( x-y\right) ,  \label{eq 3.4}
\end{align}%
as presented to the scalar, fermionic and gauge fields, respectively. Moreover, by the locality postulate \cite{10}, the support of the causal propagator is given by \eqref{eq 4.2a}. We shall now deduce them explicitly for the three stated cases.

\subsection{Scalar commutator}

Recalling the previous result \eqref{eq 2.4} for the scalar field, we have that the scalar causal propagator is given by:%
\begin{equation}
\hat{D}_{m}\left( p\right) =\frac{i}{2\pi }sgn\left( p^{-}\right) \delta
\left( 2p^{+}p^{-}-\omega _{m}^{2}\right) .  \label{eq 3.5}
\end{equation}%
Now, to find this commutator in the configuration space, we should calculate the following Fourier transformation%
\begin{equation}
D_{m}\left( x\right) =i\left( 2\pi \right) ^{-3}\int d^{2}k_{\bot
}dk^{+}dk^{-}sgn\left( k^{-}\right) \delta \left( 2k^{-}k^{+}-\omega
_{m}^{2}\right) e^{-i\left( k^{-}x^{+}+k^{+}x^{-}+k_{\bot }x^{\bot }\right)}.
\end{equation}%
We can perform the integration in $k^{-}$ by making use of the $\delta $-function,%
\begin{equation}
D_{m}\left( x\right) =i\left( 2\pi \right) ^{-3}\frac{1}{2}\int d^{2}k_{\bot
}\int dk^{+}\frac{1}{k^{+}}e^{-i\left( \frac{\omega _{m}^{2}}{2k^{+}}%
x^{+}+k^{+}x^{-}+k_{\bot }x^{\bot }\right) }.
\end{equation}%
Besides, identifying the differential and inverse-differential operators: $k_{\bot }^{2}\rightarrow \left( i\partial _{\bot }\right) ^{2}$, $\left( k^{+}\right)^{-1}  \rightarrow \left(i\partial _{-}\right)^{-1}$, it follows
\begin{align}
D_{m}\left( x\right) &= \frac{1}{2\partial _{-}}e^{-i\left( \frac{\left(
i\partial _{\bot }\right) ^{2}+m^{2}}{2i\partial _{-}}x^{+}\right) }\left(
2\pi \right) ^{-3}\int d^{2}k_{\bot }\int dk^{+}e^{-i\left(
k^{+}x^{-}+k_{\bot }x^{\bot }\right) },  \notag \\
&= \frac{1}{2\partial _{-}}e^{-\left( \frac{\left( i\partial _{\bot }\right)
^{2}+m^{2}}{2\partial _{-}}x^{+}\right) }\delta \left( x^{-}\right) \delta
\left( x^{\bot }\right) ,
\end{align}%
in the second equality we have identified the temporal and the two-dimensional transverse $\delta $-distributions. Moreover, after some algebraic manipulation, one gets%
\begin{equation}
D_{m}\left( x\right) =\frac{1}{2}\sum\limits_{n=0}^{\infty }\frac{\left(
-x^{+}\right) ^{n}}{2^{n}n!}\left( \partial _{-}\right) ^{-\left( n+1\right)
}\delta \left( x^{-}\right) \left[ \left( i\partial _{\bot }\right)
^{2}+m^{2}\right] ^{n}\delta \left( x^{\bot }\right) . \label{eq 3.34}
\end{equation}%
Finally, by using the distributional identity $\left( \partial \right) ^{-n}\delta \left(x\right) =\frac{1}{2}sgn\left( x\right) \frac{x^{n-1}}{\left( n-1\right) !}$, $n\geq 1$, Eq.\eqref{b.1}, we obtain
\begin{equation}
D_{m}\left( x\right) =\frac{1}{4}sgn\left( x^{-}\right)
\sum\limits_{n=0}^{\infty }\frac{\left( -x^{+}x^{-}\right) ^{n}}{2^{n}\left(
n!\right) ^{2}}\left[ \left( i\partial _{\bot }\right) ^{2}+m^{2}\right]
^{n}\delta \left( x^{\bot }\right) .  \label{eq 3.6}
\end{equation}%
From the expression \eqref{eq 3.6} we see that the propagator is symmetric under $x^{+}\leftrightarrow x^{-}$, since $sgn\left( x^{+}\right) =sgn\left(x^{-}\right) $. Therefore, since it obeys $x^{+}x^{-}\geq 0$ and that the support of $%
\delta \left( x^{\bot }\right) $ and its derivatives is contained in $x^{\bot }=0^{\bot }$, we can state that the support of $D_{m}\left( x\right) $ is contained in the region where $x^{2}\geq 0$, then this shows that this distribution have causal support.

Another important aspect to analyse is the series convergence, one may notice that the series converges only if $-x^{+}x^{-}\leq 0$, otherwise this is not a well-defined distribution. Hence, we can write the expression \eqref{eq 3.6} in terms of known functions,%
\begin{equation}
D_{m}\left( x\right) =\frac{sgn\left( x^{-}\right) }{2\pi }\left[ \delta
\left( x^{2}\right) -\frac{m}{2}\frac{\theta \left( x^{2}\right) }{\sqrt{%
x^{2}}}J_{1}\left( m\sqrt{x^{2}}\right) \right] .  \label{eq 3.8}
\end{equation}%
This result looks like the usual one written in instant-form coordinates \cite{12}, where $x^{0}$ is taken as the parameter responsible for the dynamical evolution of the system and $p^{0}$ as the pole parameter.

\subsubsection{ Equal-time scalar commutator}

In the instant-form coordinates the commutator is evaluated at $x^{0}=0$, which corresponds in taking this limit in \eqref{eq 3.8}; also, this implies that $x^{2}=-\vec{x}^{2}<0$ for $\vec{x}\neq \vec{0}$, then it is clear that $D_{m}\left( x\right) =0$. In this case we do not have any problem because $x $ is outside of the support of the singular distribution $\delta \left(x^{2}\right) $.

Nevertheless, in the light-front case if we take $x^{+}=0$ in \eqref{eq 3.8}, we have points as $\left( 0,0^{\bot },x^{-}\right) $ that are inside the support of $\delta \left( x^{2}\right) $. Thus, we shall take $x^{+}=0$ in the series \eqref{eq 3.6}, and it is only the term $n=0$ that survives after this limit
\begin{equation}
D_{m}\left( 0,x^{\bot },x^{-}\right) =\frac{1}{4}sgn\left( x^{-}\right)
\delta \left( x^{\bot }\right) .  \label{eq 3.9a}
\end{equation}%
With these results we find the equal-time scalar commutator \eqref{eq 3.2}%
\begin{equation}
\left[ \phi \left( x\right) ,\phi \left( y\right) \right] _{x^{+}=y^{+}}=-i%
\frac{1}{4}sgn\left( x^{-}-y^{-}\right) \delta \left( x^{\bot }-y^{\bot
}\right) .  \label{eq 3.9}
\end{equation}

\subsection{Fermionic anticommutator}

Now, for the fermionic propagator, we can recall the result \eqref{eq 2.5a} to then write the propagator in the configuration space such as%
\begin{equation}
S\left( x\right) =\left( i\gamma .\partial +m\right) D_{m}\left( x\right) ,
\label{eq 3.10}
\end{equation}%
where $D_{m}\left( x\right) $ is the scalar propagator \eqref{eq 3.8}. Since the support of a distribution also contains the support of its derivatives, then
\begin{equation}
\text{Supp}~S\left( x\right) \subset \text{Supp}~D_{m}\left( x\right) \subset \bar{
V}^{+}\left( x\right) \cup \bar{V}^{-}\left( x\right) ,
\end{equation}%
which is the principal characteristic of causal propagators. Then, by its definition \eqref{eq 3.3}, it follows that the fermionic anticommutator is given by%
\begin{equation}
\left\{ \psi \left( x\right) ,\bar{\psi}\left( y\right) \right\} =-i\left(
i\gamma .\partial +m\right) D_{m}\left( x-y\right) .  \label{eq 3.10a}
\end{equation}

\subsubsection{Equal-time fermionic anticommutator}

From the expression \eqref{eq 3.10a}, we can separate the fermionic propagator, at $x^{+}=0$, into the longitudinal, temporal, transverse and massive parts%
\begin{equation}
S\left( x\right) =\left[ \left( i\gamma .\partial ^{+}+i\gamma .\partial
^{-}+i\gamma .\partial ^{\bot }+m\right) D_{m}\left( x\right) \right]
_{x^{+}=0},  \label{eq 3.11}
\end{equation}%
respectively. We have also defined $\gamma .\partial ^{+}=\gamma ^{+}\partial _{+}$, $\gamma .\partial ^{-}=\gamma ^{-}\partial _{-}$, and $\gamma .\partial ^{\bot }=\gamma ^{\bot }\partial _{\bot }$. Into the last three terms of \eqref{eq 3.11} we may take directly the limit $x^{+}=0$ in $D_{m}\left( x\right) $, resulting into%
\begin{equation}
\left[ S^{-}+S^{\bot }+S^{m}\right] \left( 0,x^{\bot },x^{-}\right) =\left(
i\gamma .\partial ^{-}+i\gamma .\partial ^{\bot }+m\right) D_{m}\left(0,x^{\bot },x^{-}\right) .
\end{equation}%
Moreover, using the expression of $D_{m}\left( 0,x^{\bot },x^{-}\right)$ obtained in \eqref{eq 3.9a}, one gets%
\begin{equation}
\left[ S^{-}+S^{\bot }+S^{m}\right] \left( 0,x^{\bot },x^{-}\right) =\frac{i%
}{2}\left( \gamma ^{-}\right) \delta \left( x^{-}\right) \delta \left(
x^{\bot }\right) +\frac{1}{4}sgn\left( x^{-}\right) \left( i\gamma .\partial
^{\bot }+m\right) \delta \left( x^{\bot }\right) ,  \label{eq 3.12a}
\end{equation}%
where we have used the identity $\frac{1}{2}\partial _{-}sgn\left( x^{-}\right)=\delta \left( x^{-}\right) $. Besides, for the longitudinal part of \eqref{eq 3.11} we have to be cautious and use instead the series form \eqref{eq 3.6} for the scalar causal propagator
\begin{equation}
S^{+}\left( x\right) =i\gamma .\partial ^{+}D_{m}\left( x\right) =-i\gamma
^{+}\frac{1}{4}\left\vert x^{-}\right\vert \sum\limits_{n=1}^{\infty }\frac{%
\left( -x^{+}x^{-}\right) ^{n-1}}{2^{n}n!\left( n-1\right) !}\left[ \left(
i\partial _{\bot }\right) ^{2}+m^{2}\right] ^{n}\delta \left( x^{\bot
}\right) .
\end{equation}%
Now the limit $x^{+}=0$ can be taken without any complication%
\begin{equation}
S^{+}\left( 0,x^{\bot },x^{-}\right) =-\frac{i}{8}\left\vert
x^{-}\right\vert \left\{ \gamma ^{+}\left[ \left( i\gamma .\partial _{\bot
}\right) ^{2}+m^{2}\right] \right\} \delta \left( x^{\bot }\right) .
\label{eq 3.12}
\end{equation}%
Therefore, adding \eqref{eq 3.12a} and \eqref{eq 3.12}, we finally obtain the fermionic causal propagator at $x^{+}=0$%
\begin{align}
S\left( 0,x^{\bot },x^{-}\right) &= \frac{i}{2}\left( \gamma ^{-}\right)
\delta \left( x^{-}\right) \delta \left( x^{\bot }\right) +\frac{1}{4}%
sgn\left( x^{-}\right) \left( i\gamma .\partial ^{\bot }+m\right) \delta
\left( x^{\bot }\right)  \notag \\
& \quad-\frac{i}{8}\left\vert x^{-}\right\vert  \gamma ^{+}\left[ \left(
i\gamma .\partial _{\bot }\right) ^{2}+m^{2}\right] \delta \left(
x^{\bot }\right) .  \label{eq 3.13}
\end{align}
Then, the equal-time fermionic anticommutator reads
\begin{align}
\left\{ \psi \left( x\right) ,\bar{\psi}\left( y\right) \right\}
_{x^{+}=y^{+}} &= \frac{1}{2}\bigg\{\left( \gamma ^{-}\right) \delta \left(
x^{-}-y^{-}\right) -\frac{i}{2}sgn\left( x^{-}-y^{-}\right) \left( i\gamma .\partial _{x}^{\bot }+m\right) \notag \\
& \quad-\frac{1}{8}\left\vert x^{-}-y^{\bot }\right\vert \gamma ^{+}\left[ \left(
i\gamma .\partial ^{\bot }_{x}\right) ^{2}+m^{2}\right] \bigg\}\delta \left(
x^{\bot }-y^{\bot }\right)  \label{eq 3.14}
\end{align}%
This result is in agreement with the one obtained previously by canonical methods \cite{40}.

\subsection{Electromagnetic commutator}

At last, the starting point to determine the commutation relations for the gauge field components is the causal propagator \eqref{eq 2.17}%
\begin{equation}
\hat{D}_{\mu \nu }\left( k\right) =h_{\mu \nu }\hat{D}_{0}\left(
k;1,0\right) -\left( k_{\mu }\eta _{\nu }+k_{\nu }\eta _{\mu }\right) \hat{D}%
_{0}\left( k;1,1\right) +\eta _{\mu }\eta _{\nu }\hat{D}_{0}\left(
k;0,2\right) ,  \label{eq 3.15}
\end{equation}%
where we have defined the quantities Eqs.\eqref{eq 2.11}, \eqref{eq 2.14} and \eqref{eq 2.16}%
\begin{gather}
\hat{D}_{0}\left( k;1,0\right) = \frac{i}{2\pi }sgn\left( k^{-}\right)
\delta \left( 2k^{+}k^{-}-\omega _{0}^{2}\right) ,\quad \hat{D}_{0}\left( k;0,2\right) = -\frac{i}{2\pi }\delta ^{\left( 1\right)
}\left( k^{+}\right),  \\
\hat{D}_{0}\left( k;1,1\right) = \frac{i}{2\pi }\frac{\left\vert
2k^{-}\right\vert }{\omega _{0}^{2}}\left[ \delta \left( 2k^{+}k^{-}-\omega
_{0}^{2}\right) -\delta \left( 2k^{+}k^{-}\right) \right].
\end{gather}%
Furthermore, rewriting the propagator \eqref{eq 3.15} in the configuration space%
\begin{equation}
D_{\mu \nu }\left( x\right) =h_{\mu \nu }D_{0}\left( x;1,0\right) -i\left(
\eta _{\nu }\partial _{\mu }+\eta _{\mu }\partial _{\nu }\right) D_{0}\left(
x;1,1\right) +\eta _{\mu }\eta _{\nu }D_{0}\left( x;0,2\right).
\label{eq 3.16}
\end{equation}%
Thus, our task is now to evaluate the components of the propagator $D_{0}\left( x;n,l\right) $ for $\left( n,l\right) =\left( 1,0\right) ;\left( 1,1\right) ;\left( 0,2\right) $.

\begin{enumerate}[label=\textbf{\roman*}]

\item For $\left( n,l\right) =\left( 1,0\right) $, we see that $\hat{D}_{0}\left( k;1,0\right) =\hat{D}_{0}\left( k\right) $, is the massless scalar causal propagator. Thus, taking $m=0$ in the series \eqref{eq 3.6}%
\begin{equation}
D_{0}\left( x;1,0\right) =D_{0}\left( x\right) =\frac{1}{4}sgn\left(
x^{-}\right) \sum\limits_{n=0}^{\infty }\frac{\left( -x^{+}x^{-}\right) ^{n}%
}{2^{n}\left( n!\right) ^{2}}\left[ \left( i\partial _{\bot }\right) ^{2}%
\right] ^{n}\delta \left( x^{\bot }\right) . \label{eq 3.18}
\end{equation}%
As it happens in the massive case the support of massless propagator $D_{0}$ is contained in the region $x^{2}\geq 0$, then this distribution has causal support. On the other hand, it should be stressed that the series convergence holds if $-x^{+}x^{-}\leq 0$. Moreover, to find its convergence range we can take the limit $m\rightarrow 0^{+}$ in \eqref{eq 3.8}%
\begin{equation}
D_{0}\left( x;1,0\right) =\lim_{m\rightarrow 0^{+}}D_{m}\left( x\right) =%
\frac{sgn\left( x^{+}\right) }{2\pi }\delta \left( x^{2}\right) .
\end{equation}

\item For $\left( n,l\right) =\left( 1,1\right) $ we have to evaluate the following Fourier transformation%
\begin{align}
D_{0}\left( x;1,1\right) &= i\left( 2\pi \right) ^{-3}\int d^{2}k_{\bot
}dk^{+}dk^{-}\frac{\left\vert 2k^{-}\right\vert }{\omega _{0}^{2}}\notag \\
& \quad \times\left[\delta \left( 2k^{+}k^{-}-\omega _{0}^{2}\right) -\delta \left(
2k^{+}k^{-}\right) \right] e^{-i\left( k^{-}x^{+}+k^{+}x^{-}+k_{\bot
}x^{\bot }\right) },  \notag \\
&= i\left( 2\pi \right) ^{-3}\int d^{2}k_{\bot }dk^{-}\frac{1}{\omega
_{0}^{2}}\left[ e^{-i\frac{\omega _{0}^{2}}{2k^{-}}x^{-}}-1\right]
e^{-i\left( k^{-}x^{+}+k_{\bot }x^{\bot }\right) }.
\end{align}%
In order to evaluate this expression we may identify the differential and inverse-differential operators $k_{\bot }^{2}\rightarrow \left( i\partial _{\bot }\right) ^{2}$, $\left( k_{\bot }^{2}\right)^{-1}  \rightarrow \left[ \left(i\partial _{\bot }\right)^{2}\right]^{-1} $, $\left( k^{-}\right)^{-1}  \rightarrow \left(i\partial _{+}\right)^{-1}$,
\begin{align}
D_{0}\left( x;1,1\right) &= i\frac{1}{\left( i\partial _{\bot }\right) ^{2}}%
\left[ e^{-\frac{\left( i\partial _{\bot }\right) ^{2}}{2\partial _{+}}%
x^{-}}-1\right] \delta \left( x^{+}\right) \delta \left( x^{\bot }\right) ,
\notag \\
&= i\frac{1}{\left( i\partial _{\bot }\right) ^{2}}\sum\limits_{n=1}^{\infty
}\frac{1}{n!}\left[ -\frac{\left( i\partial _{\bot }\right) ^{2}}{2\partial
_{+}}x^{-}\right] ^{n}\delta \left( x^{+}\right) \delta \left( x^{\bot
}\right) , \label{3.35}
\end{align}%
in which we have identified the longitudinal and the two dimensional transverse $\delta$-distributions. Moreover, it is convenient to separate the first term of the sum in such a way%
\begin{equation}
D_{0}\left( x;1,1\right) =-\frac{i}{2}x^{-}\left( \partial _{+}\right)
^{-1}\delta \left( x^{+}\right) \delta \left( x^{\bot }\right)
+i\sum\limits_{n=2}^{\infty }\frac{\left( -x^{-}\right) ^{n}\left( \partial
_{+}\right) ^{-n}\delta \left( x^{+}\right) }{2^{n}n!}\left[ \left(
i\partial _{\bot }\right) ^{2}\right] ^{\left( n-1\right) }\delta \left(
x^{\bot }\right) ,
\end{equation}%
and, by using the identity $\left( \partial \right) ^{-n}\delta \left(x\right) =\frac{1}{2}sgn\left( x\right) \frac{x^{n-1}}{\left( n-1\right) !}$, Eq.\eqref{b.1}, one may rewrite the second term as the following suitable form%
\begin{equation}
D_{0}\left( x;1,1\right) =-\frac{i}{4}\left\vert x^{-}\right\vert \delta
\left( x^{\bot }\right) -i\frac{1}{2}x^{-}sgn\left( x^{+}\right)
\sum\limits_{n=2}^{\infty }\frac{\left( -x^{+}x^{-}\right) ^{n-1}}{%
2^{n}n!\left( n-1\right) !}\left[ \left( i\partial _{\bot }\right) ^{2}%
\right] ^{\left( n-1\right) }\delta \left( x^{\bot }\right) .
\label{eq 3.20}
\end{equation}%
At first sight we can not guarantee that the first term has causal support, but this can be determined for the whole distribution. We can note that the last term is related to $D_{0}$ \eqref{eq 3.18}; therefore, the convergence holds
for $x^{+}x^{-}\geq 0$ and since the support of $\delta \left( x^{\bot }\right) $ and of its derivatives is contained in $x^{\bot }=0^{\bot }$, then $D_{0}\left( x;1,1\right) $ has causal support.

\item For $\left( n,l\right) =\left( 0,2\right) $ we have to evaluate the integrals%
\begin{align}
D_{0}\left( x;0,2\right) &= -i\left( 2\pi \right) ^{-3}\int d^{2}k_{\bot
}dk^{+}dk^{-}\delta ^{\left( 1\right) }\left( k^{+}\right) e^{-i\left(
k^{-}x^{+}+k^{+}x^{-}+k_{\bot }x^{\bot }\right) }  \notag \\
&= x^{-}\left( 2\pi \right) ^{-3}\int d^{2}k_{\bot }dk^{-}e^{-i\left(
k^{-}x^{+}+k_{\bot }x^{\bot }\right) }.
\end{align}%
Finally identifying the longitudinal and the two dimensional transverse distributions%
\begin{equation}
D_{0}\left( x;0,2\right) =x^{-}\delta \left( x^{+}\right) \delta \left(
x^{\bot }\right).  \label{eq 3.19}
\end{equation}

We can see that any point support have the form $\left( 0,0^{\bot },x^{-}\right) $ then for this points $x^{2}=0$; hence, $D_{0}\left(x;0,2\right) $ have causal support.
\end{enumerate}

With this discussion we have demonstrated that all these distributions have in fact causal support, showing thus that the electromagnetic causal propagator \eqref{eq 3.16} has causal support as well. Now we evaluate explicitly each one of the nonvanishing components from the gauge field propagator \eqref{eq 3.16}. First, we may notice that since we have considered the vector $ \eta^{\mu}=(0,0,0,1) $ and the metric tensor \eqref{A.0}, one can show from \eqref{eq 3.16} that the components $ D_{r-}\left( x\right) $ and $ D_{--}\left( x\right) $ are in fact vanishing. Besides, for the longitudinal-temporal mixed part:
\begin{equation}
D_{+-}\left( x\right)=D_{0}\left( x\right)-i\partial _{-}D_{0}\left( x;1,1\right) .  \label{eq 3.33}
\end{equation}%
Moreover, taking $ m=0 $ in \eqref{eq 3.34} and considering its symmetric property by the interchange of the variables $ x^{+}\rightleftharpoons x^{-} $, we obtain
\begin{equation}
D_{0}\left( x\right) =\frac{1}{2\partial_{+}}\sum\limits_{n=0}^{\infty
}\frac{1}{n!}\left[ -\frac{\left( i\partial _{\bot }\right) ^{2}}{2\partial
_{+}}x^{-}\right] ^{n}\delta \left( x^{+}\right) \delta \left( x^{\bot }\right) .
\end{equation}%
Then, we can see that this expression is exactly equal to $ i\partial _{-}D_{0}\left( x;1,1\right) $, Eq.~\eqref{3.35}. Hence, we have that 
\begin{equation}
D_{+-}\left( x\right)=0 ,  \label{eq 3.36}
\end{equation}%
and, since the electromagnetic propagator is a symmetric tensor, we also have that $ D_{-+}\left( x\right)=0 $. Moreover, from \eqref{eq 3.16} it is easily seen that the \emph{transverse} part is given by%
\begin{equation}
D_{rs}\left( x\right) =h_{rs}D_{0}\left( x;1,0\right) =h_{rs}D_{0}\left(
x\right) .  \label{eq 3.22}
\end{equation}%
Since $D_{0}\left( x\right) $ has causal support then it follows that $D_{rs}\left(x\right) $ has as well. Besides, the \emph{transverse-longitudinal mixed} part from \eqref{eq 3.16} is given by%
\begin{align}
D_{r+}\left( x\right) &= -i\partial _{r}D_{0}\left( x;1,1\right)  \notag \\
&= -\frac{\left\vert x^{-}\right\vert }{4}\partial _{r}\delta \left( x^{\bot
}\right) -\frac{x^{-}}{2}sgn\left( x^{+}\right) \sum\limits_{n=2}^{\infty }%
\frac{\left( -x^{+}x^{-}\right) ^{n-1}}{2^{n}n!\left( n-1\right) !}\partial
_{r}\left[ \left( i\partial _{\bot }\right) ^{2}\right] ^{\left( n-1\right)
}\delta \left( x^{\bot }\right) .  \label{eq 3.26}
\end{align}%
It follows from the causal support of $D_{0}\left( x;1,1\right) $ that the support of $D_{r+}\left( x\right) $ is also causal. Finally, the \emph{longitudinal} part from \eqref{eq 3.16} is
\begin{equation}
D_{++}\left( x\right) =-2i\partial _{+}D_{0}\left( x;1,1\right) +D_{0}\left(
x;0,2\right).
\end{equation}%
Moreover, from the expression \eqref{eq 3.20} for $D_{0}\left( x;1,1\right) $, one finds%
\begin{equation}
-2i\partial _{+}D_{0}\left( x;1,1\right) =-x^{-}\delta \left( x^{+}\right) \delta \left( x^{\bot }\right)
+\sum\limits_{n=2}^{\infty }\frac{\left( -x^{-}\right) ^{n}\left( \partial
_{+}\right) ^{-\left( n-1\right) }\delta \left( x^{+}\right) }{2^{n-1}n!}%
\left[ \left( i\partial _{\bot }\right) ^{2}\right] ^{\left( n-1\right)
}\delta \left( x^{\bot }\right) .
\end{equation}%
Hence, after some distributional manipulation one gets%
\begin{equation}
D_{++}\left( x\right) =\left( x^{-}\right) ^{2}sgn\left( x^{+}\right)
\sum\limits_{n=2}^{\infty }\frac{\left( -x^{+}x^{-}\right) ^{n-2}}{%
2^{n}n!\left( n-2\right) !}\left[ \left( i\partial _{\bot }\right) ^{2}%
\right] ^{\left( n-1\right) }\delta \left( x^{\bot }\right) .
\label{eq 3.27}
\end{equation}%
Similar to the analysis of $D_{0}$, the convergence of the component $D_{++}$ is guaranteed for $x^{+}x^{-}\geq 0$, and as the support of the derivatives of $\delta \left(x^{\bot }\right) $ is contained in $x^{\bot }=0^{\bot }$, then we have that this propagator also has causal support. So far, we have shown that any temporal and temporal mixed components of the electromagnetic propagator are identically vanishing. Moreover, we will show next which are the physical components of the field by evaluating explicitly the commutation relations at equal-time.

\subsubsection{Equal-time electromagnetic commutator}

We can evaluate the nonvanishing commutation relations among the physical components based on \eqref{eq 3.4} and the results for the propagator Eqs.\eqref{eq 3.22}, \eqref{eq 3.26} and \eqref{eq 3.27}. Therefore, taking the limit $x^{+}=y^{+}$ we may obtain the equal-time commutators. First, for the \emph{transverse} components one finds%
\begin{equation}
\left[ A_{r}\left( x\right) ,A_{s}\left( y\right) \right] _{x^{+}=y^{+}}=i%
\frac{h_{rs}}{4}sgn\left( x^{-}-y^{-}\right) \delta \left( x^{\bot }-y^{\bot
}\right) ,  \label{eq 3.28}
\end{equation}%
whereas, for the \emph{transverse-longitudinal mixed} components, it follows that the commutation relation is%
\begin{equation}
\left[ A_{r}\left( x\right) ,A_{+}\left( y\right) \right] _{x^{+}=y^{+}}=-i%
\frac{1}{4}\left\vert x^{-}-y^{-}\right\vert \partial _{r}\delta \left(
x^{\bot }-y^{\bot }\right) .  \label{eq 3.31}
\end{equation}%
At last, the equal-time commutation relation for the \emph{longitudinal} components reads as%
\begin{equation}
\left[ A_{+}\left( x\right) ,A_{+}\left( y\right) \right] _{x^{+}=y^{+}}=i%
\frac{1}{8}%
\left( x^{-}-y^{-}\right) ^{2}sgn\left( x^{-}-y^{-}\right) \left[ \left(
i\partial _{\bot }\right) ^{2}\right] \delta \left( x^{\bot }-y^{\bot
}\right) .  \label{eq 3.32}
\end{equation}%
One can easily show, in the distributional sense, that the result \eqref{eq 3.28} is associated to positive norm states, whereas Eqs.\eqref{eq 3.31} and \eqref{eq 3.32} do not have defined sign. Therefore, it follows that only the transverse components are the physical components of the gauge field. We hope that, with the development and results of the last two sections, we have shown how powerful and simple the axiomatic approach is (with no inconsistency neither misleading concepts), once it makes use only of physical concepts, such as causality, in its development and subsequent outcome. In our case we have constructed the commutation relations for the physical components of the fields, by following a well-defined distributional approach.


\section{Causal method}

\label{sec:4}

After having showed the strength of the axiomatic approach, when the distributional character of the fields are taken into account, leading therefore to well-defined outcome, we are now in position to proceed to our main purpose here, which relies on the discussion upon the light-cone poles of the type $1/\left( k^{+}\right) ^{n}$. As discussed earlier, this problem arises because a Wick rotation is not allowed since the rotating-line crosses the poles, requiring thus a suitable prescription to deal consistently with the poles and then obtaining the correct Feynman integrals. Our main aim now is to show, without requiring or by making use of a prescription, but only regarding in an axiomatic causal theory, the so-called Epstein-Glaser causal perturbative method, how this illness from the light-front theory is not present.

In the Epstein-Glaser's causal method \cite{18}, the $S$-matrix is constructed without making any reference to the Hamiltonian formalism, its explicit form is obtained by making use of certain physical conditions -- with causality playing a major role. \footnote{A complete description and development of the causal approach can be found in our second paper \cite{49}.} For our purposes, it suffices to define two distributions in the theory. First, the general definition for the \emph{Feynman} propagator $D^{F}$, Eq.~\eqref{eq 4.6}, corresponds to a distribution that indicates the propagation of free particles in the correct time direction. Moreover, we also have defined previously the \emph{causal} propagator $D$, Eq.~\eqref{eq 4.2}, in such a way that the distribution indicates the propagation of free particles with velocity no greater than light-velocity. This distribution can be separated into two distribution, by the splitting of its support, such as%
\begin{equation}
D\left( x\right) =D^{R}\left( x\right) -D^{A}\left( x\right) ,
\label{eq 5.1}
\end{equation}%
where $D^{R}$ and $D^{A}$ are the \emph{retarded} and the \emph{advanced} propagators \eqref{eq 4.5}, respectively. In principle, we can split these propagators into a positive and negative frequency part. Thus, from \eqref{eq 5.1}, one finds%
\begin{equation}
D^{\left( \pm \right) }\left( x\right) =D^{R\left( \pm \right) }\left(
x\right) -D^{A\left( \pm \right) }\left( x\right) .  \label{eq 5.3}
\end{equation}%
Moreover, we may recall the definition \eqref{eq 4.6}, and then write the Feynman propagator also in terms of the retarded and advanced propagators%
\begin{equation}
D^{F}\left( x\right) =D^{R\left( +\right) }\left( x\right) +D^{A\left(
-\right) }\left( x\right) ,  \label{eq 5.4}
\end{equation}%
or equivalently, by recalling the relation \eqref{eq 4.9}, rewrite it as
\begin{equation}
D^{F}\left( x\right) =D^{R}\left( x\right) -D^{\left( -\right) }\left(
x\right) =D^{A}\left( x\right) +D^{\left( +\right) }\left( x\right) .  \label{eq 5.4a}
\end{equation}%
This strong relation is a result of general principles, which we can use to determine the Feynman propagator. In the previous Sect. \ref{sec:1} and \ref{sec:2} we have determined a general formula to find the positive and negative frequency propagators, Eq.~\eqref{eq 1.21}. However, in the Epstein-Glaser's approach, the splitting of the causal propagator into its advanced and retarded part is in fact more laborious \cite{49}, and it follows a set of well-defined rules as it will be shown now.

In general, the operator-valued distributions which we shall have to split are written in the form%
\begin{equation}
D_{n}\left( x_{1},...,x_{n}\right) =\underset{k}{\sum }:\underset{j}{\prod }%
\varphi ^{\dag }\left( x_{j}\right) d_{n}^{k}\left( x_{1},...,x_{n}\right)
\underset{l}{\prod }\varphi \left( x_{l}\right) \underset{m}{\prod }A\left(
x_{m}\right) :,  \label{eq 5.2}
\end{equation}%
where $\varphi $, $\varphi ^{\dag }$ are free charged (bosons or fermions) matter fields and $A$ stands for the free gauge fields. In this expression $d_{n}^{k}$ are numerical tempered distributions, $d_{n}^{k}\in \mathbf{S} ^{\prime }\left( \mathbb{R}^{4\mathbf{n}}\right) $, with causal support. Moreover, because of its translational invariance, it is sufficient to put $x_{n}=0$ and consider%
\begin{equation}
d\left( x\right) \equiv d_{n}^{k}\left( x_{1},...,x_{n-1},0\right) \in
\mathbf{S}^{\prime }\left( \mathbb{R}^{\mathbf{m}}\right) ,\quad \mathbf{m}=4%
\mathbf{n}-4.  \label{eq 5.0}
\end{equation}%
As discussed above, a rather nontrivial step is the splitting of the numerical causal distribution $d$ into the (numerical) advanced and retarded distributions $a$ and $r$, respectively. When we analyse the convergence of the sequence $\left\{ \left\langle d,\phi _{\alpha }\right\rangle \right\} $, where $\phi_{\alpha }$ has decreasing support when $\alpha \rightarrow 0^{+}$ and belongs to the Schwartz space $\mathbf{S}$, it follows that $d$ is called \textit{singular} of order $\omega $ if its Fourier transform $\hat{d}\left( p\right) $ has a quasi-asymptotic $\hat{d}_{0}\left( p\right) \neq 0$ at $p=\infty $ with regard to a positive continuous function $\rho \left( \alpha \right) $, $\alpha >0$, i.e., if the limit%
\begin{equation}
\lim_{\alpha \rightarrow 0^{+}}\rho \left( \alpha \right) \left\langle \hat{d%
}\left( \frac{p}{\alpha }\right) ,\phi \left( p\right) \right\rangle
=\left\langle \hat{d}_{0}\left( p\right) ,\phi \left( p\right) \right\rangle
\neq 0,  \label{eq 5.6}
\end{equation}%
exists in $\mathbf{S}^{\prime }\left( \mathbb{R}^{m}\right) $, with the \textit{power-counting} function $\rho \left( \alpha \right) $ satisfying
\begin{equation}
\lim_{\alpha \rightarrow 0}\frac{\rho \left( a\alpha \right) }{\rho \left(
\alpha \right) }=a^{\omega },\quad \forall ~ a>0,  \label{eq 5.8}
\end{equation}%
with%
\begin{equation}
\rho \left( \alpha \right) \rightarrow \alpha ^{\omega }L\left( \alpha
\right) ,\text{ when }\alpha \rightarrow 0^{+},  \label{eq 5.7}
\end{equation}%
where $L\left( \alpha \right) $ is a quasi-constant function at $\alpha =0$. But, of course, there is an equivalent definition in the coordinate space, however, since the splitting process is more easily accomplished in the momentum space, this one suffices to our purposes. From this definition we have two distinct cases which depend on the value of $\omega $, these are \cite{49}:

\textit{(i) }\emph{Regular} distributions - for $\omega <0$, in this case the solution of the splitting problem is unique and the retarded distribution is defined by multiplying $d$ by step functions%
\begin{equation}
\hat{r}\left( p\right) =\frac{i}{2\pi }sgn\left( p_{\chi }\right)
\int_{-\infty }^{+\infty }dt\frac{\hat{d}\left( tp\right) }{\left(
1-t+sgn\left( p_{\chi }\right) i0^{+}\right) } . \label{eq 5.5}
\end{equation}

\textit{(ii) }\emph{Singular} distributions - for $\omega \geq 0$, then the solution can not be obtained as in the \emph{regular} case and, after a careful mathematical treatment \cite{49}, it may be shown that the retarded distribution is given by the central splitting solution%
\begin{equation}
\hat{r}\left( p\right) =\frac{i}{2\pi }sgn\left( p_{\chi }\right)\int_{-\infty }^{+\infty }dt\frac{%
\hat{d}\left( tp\right) }{\left( t-sgn\left( p_{\chi }\right)
i0^{+}\right) ^{\omega +1}\left( 1-t+sgn\left( p_{\chi }\right) i0^{+}\right) }.  \label{eq 5.5a}
\end{equation}
These solutions have the very important feature that they preserve the original symmetries of the theory, for instance the Lorentz covariance and gauge invariance. We must remark that $p_{\chi }\in \bar{V} ^{+}\cup \bar{V}^{-}$ is the \emph{parameter} used to to split the causal distribution into its retarded and advanced part; moreover, $p_{\chi }$ may be a time-like or light-like parameter. \footnote{We must not confuse $p_{\chi }$ with the parameter $k_{\lambda }$ in the Eq.~\eqref{eq 1.21}, which is used to split the causal propagator into its positive and negative frequency part.} In the light-front it is possible to choose either $p_{\chi }=p^{+}$ or $p_{\chi }=p^{-}$, or even in the instant-form is taken as $p_{\chi }=p^{0}$. However, as it will be shown next, all the distributions that we are dealing with here will be regular ones, then requiring the use of the solution \eqref{eq 5.5}.

In order to give a proper glance on the causal approach functionality, let us consider the propagator of the scalar fields given by Eq.~\eqref{eq 3.5}
\begin{equation}
\hat{D}_{m}\left( k\right) =\frac{i}{2\pi }sgn\left( k^{-}\right) \delta
\left( 2k^{+}k^{-}-\omega _{m}^{2}\right) .
\end{equation}%
The very first thing to do in the approach we must evaluate $\hat{D}_{m}\left(\frac{k}{\alpha }\right) $ when $\alpha \rightarrow 0^{+}$%
\begin{equation}
\hat{D}_{m}\left( \frac{k}{\alpha }\right) =\alpha ^{2}\frac{i}{2\pi }%
sgn\left( k^{-}\right) \delta \left( k^{2}-\alpha ^{2}m^{2}\right)
\rightarrow \alpha ^{2}\hat{D}_{0}\left( k\right) .  \label{eq 5.18}
\end{equation}%
From that it follows that the singular order of the causal scalar propagator $\hat{D}_{m}$ is $\omega =-2$. Therefore, we may say that $\hat{D}_{m}$ is a \emph{regular} distribution of order $-2$. The next step is now to use a correct distributional splitting of the causal propagator, which is given by the expression \eqref{eq 5.5} for $k_{\chi}=k^{-}$. Thus, we can determine the \emph{retarded} propagator,
\begin{equation}
\hat{D}_{m}^{R}\left( k\right) =\frac{i}{2\pi }sgn\left( k^{-}\right)
\int\limits_{-\infty }^{\infty }dt\frac{\hat{D}_{m}\left( tk\right) }{%
1-t+sgn\left( k^{-}\right) i0^{+}},  \label{eq 5.19}
\end{equation}%
and, by making use of the explicit expression for the propagator $\hat{D}_{m}$,
\eqref{eq 3.5}%
\begin{equation}
\hat{D}_{m}^{R}\left( k\right) =-\left( 2\pi \right) ^{-2}sgn\left(
k^{-}\right) \int\limits_{-\infty }^{\infty }dt\frac{sgn\left( tk^{-}\right)
\delta \left( t^{2}k^{2}-m^{2}\right) }{1-t+sgn\left( k^{-}\right) i0^{+}},
\end{equation}%
now, introducing the variable $s=t^{2}$, we have that%
\begin{equation}
\hat{D}_{m}^{R}\left( k\right) =-\left( 2\pi \right) ^{-2}\frac{1}{k^{2}}%
\int\limits_{0}^{\infty }ds\frac{\delta \left( s-\frac{m^{2}}{k^{2}}\right)
}{1-s+sgn\left( k^{-}\right) i0^{+}}.  \label{eq 5.20}
\end{equation}%
Therefore, we obtain the known result, defined in the proper contour,%
\begin{equation}
\hat{D}_{m}^{R}\left( k\right) =-\left( 2\pi \right) ^{-2}\frac{1}{%
k^{2}-m^{2}+sgn\left( k^{-}\right) i0^{+}} , \label{eq 5.21}
\end{equation}%
where, by condition, we have $k^{2}>0$. Now, we proceed in evaluating the cases of the fermionic and gauge fields, and then to conclude by analysing our main interest here which is the propagator associated to $1/\left( k^{+}\right) ^{n}$.


\subsection{Fermionic propagator}

For a more interesting case we consider the fermionic fields which were discussed previously in the Sect.\ref{sec:3}, when we evaluated its propagator and equal-time anticommutation relation.  \footnote{Further details on the fermionic propagator can be found in \cite{49}, where a detailed discussion on the so-called \textit{instantaneous} part of the propagator is also presented.} As obtained earlier, we have that the propagator of positive and negative frequency are given, in the momentum space, by the relations%
\begin{equation}
\hat{S}^{\left( \pm \right) }\left( k\right) =\left( \gamma .p+m\right) \hat{%
D}_{m}^{\left( \pm \right) }\left( p\right),  \label{eq 6.2}
\end{equation}%
where%
\begin{equation}
\hat{D}_{m}^{\left( \pm \right) }\left( p\right) =\pm \frac{i}{2\pi }\theta
\left( \pm p^{-}\right) \delta \left( 2p^{+}p^{-}-\omega _{m}^{2}\right) .
\label{eq 6.3}
\end{equation}%
From this result we can find the causal propagator%
\begin{equation}
\hat{S}\left( p\right) =\left( \gamma .p+m\right) \hat{D}_{m}\left( p\right),
\end{equation}%
where $\hat{D}_{m}=\hat{D}_{m}^{\left( +\right) }+\hat{D}_{m}^{\left(-\right) }$. Since $\text{Supp}~ \left(\partial ^{a}D_{m}\right)\subset \text{Supp}~D_{m}$ then the distribution $S\left( x\right) $ has causal support. The next step of the causal approach consists in determining the singular order of this distribution. For that, we must calculate $\hat{S}\left( \frac{p}{\alpha }\right)$ when $\alpha \rightarrow 0^{+}$%
\begin{equation}
\hat{S}\left( \frac{p}{\alpha }\right) =\left( \frac{\gamma .p}{\alpha }%
+m\right) \hat{D}_{m}\left( \frac{p}{\alpha }\right) \rightarrow \left(
\alpha \gamma .p+\alpha ^{2}m\right) \hat{D}_{0}\left( p\right) ,
\label{eq 6.5}
\end{equation}%
the second equality came from the scalar case \eqref{eq 5.18}. Thus, we can say that the whole propagator $S$ (or $\hat{S}$) is a \emph{regular} distribution of order $-1$. In order to evaluate the retarded distribution, we should use the splitting formula \eqref{eq 5.5} by choosing $p_{\chi }=p^{-}$. It follows then%
\begin{equation}
\hat{S}^{R}\left( p\right) =\gamma .p\frac{i}{2\pi }sgn\left( p^{-}\right)
\int\limits_{-\infty }^{+\infty }dt\frac{t\hat{D}_{m}\left( tp\right) }{%
1-t+sgn\left( p^{-}\right) i0^{+}}+m\frac{i}{2\pi }sgn\left( p^{-}\right)
\int\limits_{-\infty }^{+\infty }dt\frac{\hat{D}_{m}\left( tp\right) }{%
1-t+sgn\left( p^{-}\right) i0^{+}},  \label{eq 6.7}
\end{equation}%
moreover, one may obtain that, after an algebraic manipulation and by making use of the expression \eqref{eq 5.19}, the fermionic \emph{retarded} propagator is
\begin{equation}
\hat{S}^{R}\left( p\right) =\left( \gamma .p+m\right) \hat{D}_{m}^{R}\left(
p\right) ,  \label{eq 6.9}
\end{equation}%
where the explicit expression for $\hat{D}_{m}^{R}$ is given by Eq.~\eqref{eq 5.21}. Furthermore, one may recall the definition \eqref{eq 5.4a} and make use of the results \eqref{eq 6.2} and \eqref{eq 6.9}, to obtain the fermionic \emph{Feynman} propagator%
\begin{equation}
\hat{S}^{F}\left( p\right) =\left( \gamma .p+m\right) \hat{D}_{m}^{F}\left(
p\right) ,  \label{eq 6.10}
\end{equation}%
where
\begin{equation}
\hat{D}_{m}^{F}\left( p\right) =-\left( 2\pi \right) ^{-2}\frac{1}{%
p^{2}-m^{2}+i0^{+}},\quad p^{2}>0. \label{eq 6.11}
\end{equation}

\subsection{Electromagnetic propagator}

Finally, we discuss the electromagnetic propagator in the light-front gauge within the realm of the causal approach. We will show that when one makes use of the systematic rules of the Epstein-Glaser's causal approach there is no need in employing any prescription to deal with the propagator's poles. For this case we already have determined the expression of the causal propagator
in \eqref{eq 2.17}%
\begin{equation}
\hat{D}_{\mu \nu }\left( k\right) =h_{\mu \nu }\hat{D}_{0}\left(
k;1,0\right) -\left( k_{\mu }\eta _{\nu }+k_{\nu }\eta _{\mu }\right) \hat{D}%
_{0}\left( k;1,1\right) +\eta _{\mu }\eta _{\nu }\hat{D}_{0}\left(
k;0,2\right) ,  \label{eq 7.3}
\end{equation}%
where $\eta $ is a light-like vector with components $\eta ^{\mu }=\left(0,0,0,1\right) $, and the quantities are defined by the Eqs.\eqref{eq 2.11}, \eqref{eq 2.14} and \eqref{eq 2.16}%
\begin{gather}
\hat{D}_{0}\left( k;1,0\right) = \frac{i}{2\pi }sgn\left( k^{-}\right) \delta \left( 2k^{+}k^{-}-\omega _{0}^{2}\right) , \quad \hat{D}_{0}\left( k;0,2\right) = -\frac{i}{2\pi }\delta ^{\left( 1\right) }\left( k^{+}\right) ,  \\
 \hat{D}_{0}\left( k;1,1\right) = \frac{i}{\pi }\frac{\left\vert
k^{-}\right\vert }{\omega _{0}^{2}}\left[ \delta \left( 2k^{+}k^{-}-\omega
_{0}^{2}\right) -\delta \left( 2k^{+}k^{-}\right) \right],
\end{gather}%
and $\omega _{0}=\sqrt{k_{\bot }^{2}}$. Next, we have to verify the singular order of the whole distribution, for that we must calculate $\hat{D}_{\mu \nu }\left( \frac{k}{\alpha }\right) $ from \eqref{eq 7.3} when $\alpha\rightarrow 0^{+}$%
\begin{equation}
\hat{D}_{\mu \nu }\left( \frac{k}{\alpha }\right) =h_{\mu \nu }\hat{D}%
_{0}\left( \frac{k}{\alpha };1,0\right) -\frac{\left( k_{\mu }\eta _{\nu
}+k_{\nu }\eta _{\mu }\right) }{\alpha }\hat{D}_{0}\left( \frac{k}{\alpha }%
;1,1\right) +\eta _{\mu }\eta _{\nu }\hat{D}_{0}\left( \frac{k}{\alpha }%
;0,2\right) ,  \label{eq 7.7}
\end{equation}%
in which it follows, by using the explicit expression of each one of the terms, that%
\begin{gather}
\hat{D}_{0}\left( \frac{k}{\alpha };1,0\right) = \alpha ^{2}\hat{D}%
_{0}\left( k;1,0\right) , \quad \hat{D}_{0}\left( \frac{k}{\alpha };0,2\right) = \alpha ^{2}\hat{D}%
_{0}\left( k;0,2\right), \\
\hat{D}_{0}\left( \frac{k}{\alpha };1,1\right) = \alpha ^{3}\hat{D}%
_{0}\left( k;1,1\right)  .
\end{gather}%
Thus, the expression of the causal distribution $\hat{D}_{\mu \nu }$ reads
\begin{equation}
\hat{D}_{\mu \nu }\left( \frac{k}{\alpha }\right) =\alpha ^{2}\left[ h_{\mu
\nu }\hat{D}_{0}\left( k;1,0\right) -\left( k_{\mu }\eta _{\nu }+k_{\nu
}\eta _{\mu }\right) \hat{D}_{0}\left( k;1,1\right) +\eta _{\mu }\eta _{\nu }%
\hat{D}_{0}\left( k;0,2\right) \right] .  \label{eq 7.8}
\end{equation}%
This means that the causal propagator $\hat{D}_{\mu \nu }$ is a \emph{regular} distribution of order: $-2$ . Now we are in position to find the retarded distribution. Thus, using the regular splitting formula \eqref{eq 
5.5} with $k_{\chi }=k^{-}$, it follows%
\begin{align}
\hat{D}_{\mu \nu }^{R}\left( k\right) &= h_{\mu \nu }\frac{i}{2\pi }%
sgn\left( k^{-}\right) \int\limits_{-\infty }^{+\infty }dt\frac{\hat{D}%
_{0}\left( tk;1,0\right) }{1-t+sgn\left( k^{-}\right) i0^{+}} +\eta _{\mu }\eta _{\nu }\frac{i}{2\pi }sgn\left( k^{-}\right)
\int\limits_{-\infty }^{+\infty }dt\frac{\hat{D}_{0}\left( tk;0,2\right) }{%
1-t+sgn\left( k^{-}\right) i0^{+}} \notag \\
& \quad-\left( k_{\mu }\eta _{\nu }+k_{\nu }\eta _{\mu }\right) \frac{i}{2\pi }%
sgn\left( k^{-}\right) \int\limits_{-\infty }^{+\infty }dt\frac{t\hat{D}%
_{0}\left( tk;1,1\right) }{1-t+sgn\left( k^{-}\right) i0^{+}}  .  \label{eq 7.9}
\end{align}%
One can make use of the above explicit expression for the $\hat{D}_{0}\left(k;l,n\right) $ to then evaluate, without any complication, the dispersion integrals. It then finally follows the expression for the electromagnetic \emph{retarded} propagator%
\begin{equation}
\hat{D}_{\mu \nu }^{R}\left( k\right) =h_{\mu \nu }\hat{D}_{0}^{R}\left(
k;1,0\right) -\left( k_{\mu }\eta _{\nu }+k_{\nu }\eta _{\mu }\right) \hat{D}%
_{0}^{R}\left( k;1,1\right) +\eta _{\mu }\eta _{\nu }\hat{D}_{0}^{R}\left(
k;0,2\right) ,  \label{eq 7.13}
\end{equation}%
where
\begin{gather}
\hat{D}_{0}^{R}\left( k;1,0\right) =-\left( 2\pi \right) ^{-2}\frac{1}{%
k^{2}+sgn\left( k^{-}\right) i0^{+}}, \quad \hat{D}_{0}^{R}\left( k;0,2\right) = -\left( 2\pi \right) ^{-2}\frac{1}{%
\left( k^{+}+i0^{+}\right) ^{2}}  ,\\
\hat{D}_{0}^{R}\left( k;1,1\right) = -\left( 2\pi \right) ^{-2}\frac{1}{%
k^{+}+i0^{+}}.  \label{eq 7.14}
\end{gather}%
In order to make connection to relevant calculation it is of interest to evaluate the \emph{Feynman} propagator $\hat{D}_{\mu \nu }^{F}$. We can make use of the definition \eqref{eq 5.4a}%
\begin{equation}
\hat{D}_{\mu \nu }^{F}\left( k\right) =\hat{D}_{\mu \nu }^{R}\left( k\right)
-\hat{D}_{\mu \nu }^{\left( -\right) }\left( k\right) .  \label{eq 7.15}
\end{equation}%
Hence, from the expressions \eqref{eq 2.18} and \eqref{eq 7.13}, the \emph{Feynman} propagator reads%
\begin{equation}
\hat{D}_{\mu \nu }^{F}\left( k\right) =h_{\mu \nu }\hat{D}_{0}^{F}\left(
k;1,0\right) -\left( k_{\mu }\eta _{\nu }+k_{\nu }\eta _{\mu }\right) \hat{D}%
_{0}^{F}\left( k;1,1\right) +\eta _{\mu }\eta _{\nu }\hat{D}_{0}^{F}\left(
k;0,2\right) ,  \label{eq 7.16}
\end{equation}%
with the quantities given by%
\begin{gather}
\hat{D}_{0}^{F}\left( k;1,0\right) = -\left( 2\pi \right) ^{-2}\frac{1}{%
k^{2}+i0^{+}},  \quad \hat{D}_{0}^{F}\left( k;0,2\right) = -\left( 2\pi \right) ^{-2}\frac{1}{%
\left[ k^{+}+sgn\left( k^{-}\right) i0^{+}\right] ^{2}} ,  \\
\hat{D}_{0}^{F}\left( k;1,1\right) = -\left( 2\pi \right) ^{-2}\frac{1}{%
\omega _{0}^{2}}\left( \frac{2k^{-}}{k^{2}+i0^{+}}-\frac{1}{k^{+}+sgn\left(
k^{-}\right) i0^{+}}\right).\label{eq 7.19}
\end{gather}%
It should be noted that the \emph{Feynman} propagators obtained above are particular cases of the general case $\hat{D}_{0}^{R}\left( k;0,n\right) $, which we will calculate in the next section. Finally, we can simplify the
expression, and thus write the propagator $\hat{D}_{\mu \nu }^{F}\left( k\right) $ as the following
\begin{equation}
\hat{D}_{\mu \nu }^{F}\left( k\right) =-\left( 2\pi \right) ^{-2}\left\{
\frac{h_{\mu \nu }}{k^{2}+i0^{+}}-\frac{k_{\mu }\eta _{\nu }+k_{\nu }\eta
_{\mu }}{\left( k^{2}+i0^{+}\right) \left[ k^{+}+sgn\left( k^{-}\right)
i0^{+}\right] }+\frac{\eta _{\mu }\eta _{\nu }}{\left[ k^{+}+sgn\left(
k^{-}\right) i0^{+}\right] ^{2}}\right\} .  \label{eq 7.23}
\end{equation}%
This is the free photon propagator in the light-front gauge. Moreover, it should be emphasized that all the poles are well-defined in the expression \eqref{eq 7.23}, in which the proper (light-front) contours are explicitly shown.


\section{Propagator associated to $\frac{1}{\left( k^{+}\right) ^{n}}$}

\label{sec:5}

As it is mainly presented in the literature, when one analyse light-front field theories, a prescription is needed in order to deal with the propagator poles in a proper way. On the other hand, we have shown here so far how to deal with the scalar, fermionic and gauge fields propagator poles in the framework of the Epstein-Glaser's causal approach, where prescriptions are not necessary and all quantities are well-defined in a distributional sense. Therefore, in a way to conclude our discussion, we shall present now an analysis for a general type of propagator in the light-front. Hence, we have that the positive and negative frequency part of the propagator associated to a pole of the type $\left( k^{+}\right) ^{-n}$, for $n=1,2,3,...$, are given by the expressions \footnote{These follow from the Eq.~\eqref{eq 1.21} for $G\left( k\right) =\left(k^{+}\right) ^{-n}$ and $k_\sigma =k^+$. A similar analysis may also be accomplished for a massive pole.}%
\begin{equation}
\hat{D}^{\left( \pm \right) }\left( k;n\right) =\frac{i}{2\pi }\theta \left(
\pm k_{\lambda}\right) \frac{\left( -1\right) ^{\left( n-1\right) }}{\left(
n-1\right) !}\delta ^{\left( n-1\right) }\left( k^{+}\right) .\label{eq 8.1}
\end{equation}%
From this result we can find the causal propagator%
\begin{equation}
\hat{D}\left( k;n\right) =\frac{i}{2\pi }\frac{\left( -1\right) ^{\left(
n-1\right) }}{\left( n-1\right) !}\delta ^{\left( n-1\right) }\left(
k^{+}\right) .  \label{eq 8.2}
\end{equation}%
We can easily see that this distribution has causal support. The next step on the analysis, before determining the Feynman propagator, consists in verifying the singular order of this distribution. Hence, we must to calculate $\hat{D}_{M}\left( \frac{k}{\alpha };n\right) $ when $\alpha \rightarrow 0^{+}$%
\begin{equation}
\hat{D}_{0}\left( \frac{k}{\alpha };n\right) =\frac{i}{2\pi }\frac{\left(
-1\right) ^{\left( n-1\right) }}{\left( n-1\right) !}\delta ^{\left(
n-1\right) }\left( \frac{k^{+}}{\alpha }\right) \rightarrow \alpha ^{n}\hat{D%
}_{0}\left( k;n\right) .  \label{eq 8.4}
\end{equation}%
Therefore, we can say that the propagator $\hat{D}_{0}\left( k;n\right) $ is a \emph{regular} distribution of order $-n$. Now, from the regular splitting formula \eqref{eq 5.5}, it follows that the retarded propagator of this general pole is given by%
\begin{align}
\hat{D}_{0}^{R}\left( k;n\right)  &= \frac{i}{2\pi }sgn\left( k_{\chi
}\right) \int\limits_{-\infty }^{+\infty }dt\frac{\hat{D}_{0}\left(
tk;0,n\right) }{1-t+sgn\left( k_{\chi }\right) i0^{+}},  \notag \\
&= -\left( 2\pi \right) ^{-2}\frac{\left( -1\right) ^{\left( n-1\right) }}{%
\left( n-1\right) !}\frac{1}{\left( k^{+}\right) ^{n}}\int\limits_{-\infty
}^{+\infty }dt\frac{\delta ^{\left( n-1\right) }\left( t\right) }{%
1-t+sgn\left( k^{+}\right) i0^{+}},  \label{eq 8.6}
\end{align}%
where we have chosen $k_{\chi }=k^{+}$. The dispersion integral may be easily solved, resulting into the following expression%
\begin{equation}
\hat{D}_{0}^{R}\left( k;n\right) =-\left( 2\pi \right) ^{-2}\frac{1}{\left(
k^{+}+i0^{+}\right) ^{n}}.  \label{eq 8.8}
\end{equation}%
Replacing the results $\eqref{eq 8.1}$ and $\eqref{eq 8.8}$ into the Eq.~$\eqref{eq 5.4a}$, we obtain the \emph{Feynman} distribution%
\begin{equation}
\hat{D}_{0}^{F}\left( k;n\right) =-\left( 2\pi \right) ^{-2}\left[ \frac{1}{%
\left( k^{+}+i0^{+}\right) ^{n}}+2\pi i\theta \left( -k_{\lambda }\right)
\frac{\left( -1\right) ^{\left( n-1\right) }}{\left( n-1\right) !}\delta
^{\left( n-1\right) }\left( k^{+}\right) \right] ,  \label{eq 8.9}
\end{equation}%
where we have considered the general case for the negative frequency propagator, i.e., $k_{\lambda }$ is a time-like or light-like parameter. Moreover, the expression \eqref{eq 8.9} can be rewritten conveniently by making use of distributional identities in the such form%
\begin{equation}
\hat{D}_{0}^{F}\left( k;n\right) =-\left( 2\pi \right) ^{-2}\left[ P\frac{1}{%
\left( k^{+}\right) ^{n}}-i\pi sgn\left( k_{\lambda }\right) \frac{\left(
-1\right) ^{\left( n-1\right) }}{\left( n-1\right) !}\delta ^{\left(
n-1\right) }\left( k^{+}\right) \right] ,  \label{eq 8.11}
\end{equation}%
or equivalently as%
\begin{equation}
\hat{D}_{0}^{F}\left( k;n\right) =-\left( 2\pi \right) ^{-2}\frac{1}{\left(
k^{+}+sgn\left( k_{\lambda }\right) i0^{+}\right) ^{n}}.  \label{eq 8.12}
\end{equation}%
We can see from $\eqref{eq 8.11}$ that for a correct definition of the expression, in a distributional sense, it is necessary that $k_{\lambda }\neq k^{+}$. From these last results we may depict a parallel with some well-known prescriptions in the light-front literature.

In particular, for $n=1$ and choosing $k_{\lambda }=k^{-}$ in $\eqref{eq 8.12} $, we arrive at the known \emph{Mandelstam-Leibbrandt} prescription \cite{4,5} 
\begin{equation}
\hat{D}_{0}^{F}\left( k;1\right) =-\left( 2\pi \right) ^{-2}\frac{1}{%
k^{+}+sgn\left( k^{-}\right) i0^{+}}.  \label{eq 8.13}
\end{equation}%
A more general result can be find for an arbitrary $n$ by choosing $k_{\lambda}=\frac{k^{+}+k^{-}}{\sqrt{2}}=k_{0}$ in $\eqref{eq 8.11}$, by this choice we arrive at the \emph{Pimentel-Suzuki} prescription \cite{7}
\begin{equation}
\hat{D}_{0}^{F}\left( k;n\right) =-\left( 2\pi \right) ^{-2}\left[ P\frac{1}{%
\left( k^{+}\right) ^{n}}-i\pi sgn\left( k_{0}\right) \frac{\left( -1\right)
^{\left( n-1\right) }}{\left( n-1\right) !}\delta ^{\left( n-1\right)
}\left( k^{+}\right) \right].  \label{eq 8.14}
\end{equation}%

\subsection{Examples: Evaluation of integrals}

By means of complementarity, we shall calculate now some relevant integrals, that appear in the evaluation of radiative corrections, in order to demonstrate the strength of our result Eq.~\eqref{eq 8.12}. Hence, we shall compute the following two basic one-loop light-cone integrals (which are related to the massless two- and three-points functions)
\begin{align}
A\left( n\right) =&\int d^{2\omega }k\frac{1}{\left( k-p\right) ^{2}\left(
k^{+}\right) ^{n}},  \label{eq 8.15} \\
B\left( n\right) =&\int d^{2\omega }k\frac{1}{k^{2}\left( k-p\right)
^{2}\left( k^{+}\right) ^{n}},  \label{eq 8.16}
\end{align}%
where we are taking by means of generality $ D=2\omega $ as the dimension of space time, i.e. the limit to four dimensions means $ \omega\rightarrow2 $. Let us start by evaluating the integral $A\left( n\right) $, Eq.~\eqref{eq 8.15}, then using the result Eq.~\eqref{eq 6.11} for the massless scalar propagator and the result Eq.~\eqref{eq 8.12} for the $ n $th-order pole at $ k^{+}=0 $, with $k_{\lambda }=k^{-}$, we obtain
\begin{equation}
A\left( n\right) =\int d^{2\omega }k\frac{1}{ \left( k-p\right)
^{2}+i0^{+} }\frac{1}{\left( k^{+}+sgn\left( k^{-}\right)
i0^{+}\right) ^{n}}.
\end{equation}%
By parametrizing the massless scalar propagator with the Schwinger parametrization:
\begin{equation}
A\left( n\right)  =\left( -i\right) \int d^{2\omega
}k\int\limits_{0}^{\infty }d\alpha e^{i\alpha \left( k-p\right) ^{2}}\frac{1%
}{\left( k^{+}+sgn\left( k^{-}\right) i0^{+}\right) ^{n}},
\end{equation}
it follows that, after some algebraic manipulation of the momentum integrals, we have
\begin{align}
A\left( n\right)  =&\left( -i\right) \int\limits_{0}^{\infty }d\alpha
e^{2i\alpha \left( p^{-}p^{+}\right) }\int d^{2\omega -2}\hat{k}e^{-i\alpha
\left( \hat{k}^{2}-2\hat{k}.\hat{p}+\hat{p}^{2}\right) }  \notag \\
&\times \int\limits_{-\infty }^{\infty }dk^{-}e^{-2i\alpha \left(
k^{-}p^{+}\right) }\int\limits_{-\infty }^{\infty }dk^{+}\frac{e^{2i\alpha k^{+}\left(
k^{-}-p^{-}\right) }}{\left(
k^{+}+sgn\left( k^{-}\right) i0^{+}\right) ^{n}}. \label{eq 8.19}
\end{align}%
Nevertheless, working out the momentum integrals, we can identify the $\hat{k}$ integral with the standard Gaussian integral, although we should notice that there are two different integral region in the variable $ k^{-} $,
\begin{equation}
A\left( n\right) =\left( -i\right) \int\limits_{0}^{\infty }d\alpha
e^{2i\alpha \left( p^{-}p^{+}\right) }\left( -\frac{i\pi }{\alpha }\right)
^{\omega -1}\left[ M_{1}^{+}\left( n\right) +M_{1}^{-}\left( n\right) \right],
\end{equation}%
where we have defined
\begin{equation}
M_{1}^{\pm}\left( n\right)  =\int\limits_{-\infty}^{\infty }dk^{-}\theta \left(\pm k^{-} \right) e^{-2i\alpha
\left( k^{-}p^{+}\right) }\int\limits_{-\infty }^{\infty }dk^{+}\frac{1}{%
\left( k^{+}\pm i0^{+}\right) ^{n}}e^{2i\alpha k^{+}\left( k^{-}-p^{-}\right) },
\end{equation}%
Finally, we can make use of the residue theorem for the $ n $th-order poles $ k^{+}=\mp i0^{+} $, and after subsequent calculation, we can show that
\begin{equation}
M_{1}^{+}\left( n\right) +M_{1}^{-}\left( n\right) =\frac{\pi }{\alpha
\left( p^{+}\right) ^{n}}\left[ e^{-2i\alpha p^{+}p^{-}}-\sum_{m=0}^{n-1}%
\frac{\left( -2i\alpha p^{+}p^{-}\right) ^{m}}{m!}\right] .
\end{equation}%
Therefore, we can write the basic one-loop integral $ A(n) $ as follows
\begin{equation}
A\left( n\right) =\frac{\left( -i\right) ^{\omega } \pi ^{\omega }}{\left( p^{+}\right) ^{n}}\int\limits_{0}^{\infty }d\alpha e^{2i\alpha p^{-}p^{+}}\frac{1}{\alpha ^{\omega }}\left[ e^{-2i\alpha
p^{+}p^{-}}-\sum_{m=0}^{n-1}\frac{\left( -2i\alpha p^{+}p^{-}\right) ^{m}}{m!}\right] . \label{I1}
\end{equation}
Let us now compute the integral $ B(n) $,
\begin{equation}
B\left( n\right) =\int d^{2\omega }k\frac{1}{\left( k^{2}+i0^{+}\right) %
\left( \left( k-p\right) ^{2}+i0^{+}\right) }\frac{1}{\left( k^{+}+sgn\left(k^{-}\right) i0^{+}\right) ^{n}},
\end{equation}%
which can be rewritten as
\begin{equation}
B\left( n\right) =-\int d^{2\omega }k\int\limits_{0}^{\infty }d\gamma d\beta
e^{i\left[ \beta \left( k-p\right) ^{2}+\gamma k^{2}\right] }\frac{1}{\left(
k^{+}+sgn\left( k^{-}\right) i0^{+}\right) ^{n}}.
\end{equation}%
As a matter of notation, we can introduce $q\equiv \left( \frac{\beta }{\gamma +\beta }\right) p$ and $\alpha \equiv \gamma +\beta $, and then arrive at
\begin{align}
B\left( n\right)  =&\left( -i\right) ^{2}\int\limits_{0}^{\infty }d\gamma
d\beta e^{i\beta p^{2}}e^{i\alpha \hat{q}^{2}}\int d^{2\omega -2}\hat{k}%
e^{-i\alpha \left( \hat{k}^{2}-2\hat{k}.\hat{q}+\hat{q}^{2}\right) }  \notag\\
&\times \int\limits_{-\infty }^{\infty }dk^{-}e^{-2i\alpha
k^{-}q^{+}}\int\limits_{-\infty }^{\infty }dk^{+}e^{2i\alpha k^{+}\left(
k^{-}-q^{-}\right) }\frac{1}{\left( k^{+}+sgn\left( k^{-}\right)i0^{+}\right) ^{n}},
\end{align}%
which has a very similar expression as Eq.~\eqref{eq 8.19}. Therefore, following the same steps as outlined above, the one-loop light-cone integral $ B(n) $ reads
\begin{equation}
B\left( n\right) =\frac{\left( -i\right) ^{\omega +1}\left( \pi \right)
^{\omega }}{\left( q^{+}\right) ^{n}}\int\limits_{0}^{\infty }d\gamma d\beta
e^{i\beta p^{2}}e^{i\alpha \hat{q}^{2}}\frac{1}{\alpha ^{\omega }}\left[
e^{-2i\alpha q^{+}q^{-}}-\sum_{m=0}^{n-1}\frac{\left( -2i\alpha
q^{+}q^{-}\right) ^{m}}{m!}\right] . \label{I2}
\end{equation}
In particular, we have that the Eqs.~\eqref{I1} and \eqref{I2} for $ n=1 $ are in agreement with known results in the literature \cite{7,6,37a}.

\section{Concluding remarks}

\label{sec:6}

In this paper we have considered light-front field theories in the framework of analytic representation and also at Epstein-Glaser causal method. Since the Dirac proposal, light-front form has been applied in many different scenarios, showing that it can drastically change the content and interpretation of a given theory. Moreover, the interest on light-front field theory has showed its appealing content and led to a rich theoretical development mostly by its economical way in displaying the relevant degrees of freedom, as well as physical and richer contribution on QCD analysis. We wanted here to revisit the problem surrounding light-front field theories which is the correct definition of the poles $\left(k^{+}\right)^{-n}$. But, instead of dealing with them by employing or suggesting new a prescription, we dealt with them in a general fashion based on the use of rigorous mathematical machinery of distributions combined with strong physical concepts, such as causality. We have focused here in studying the simplest case of free fields in the light-front and defining important subtle issues, leaving the general discussion of interacting field theory to a separate paper.

The first point developed here consisted in reviewing the Wightman's formalism, and to show how the analytic representation of a propagator is obtained when physical concepts are required. From that we obtained a general framework to evaluate the positive and negative frequency parts of the propagator, and the locality of these solutions were also proved. Such quantities allowed us to evaluate subsequently the equal-time (anti)commutation relations of the scalar, fermionic and electromagnetic fields; in particular, it was showed that by following the rigorous rules one obtains directly the commutation relations for the dynamical fields only, and the redundant fields are naturally excluded. With that, we hoped to have shown in a general fashion how to obtain unambiguous and well-defined quantities in a field theory.

Consequently, with all the information gained by discussing the equal-time (anti)commutation relations through the Wightman's formalism, we were compelled to introduce more physical content in order to treat some intriguing quantities, and that led us to introduce the Epstein-Glaser causal method. Our main purpose with that discussion were to show how the illness (the need of using a prescription) of the usual light-front theory is not present in such approach. By presenting the general set up in which the causal approach is based on, we presented then the splitting solutions of the retarded propagator into the regular and singular cases in the light-front form. Subsequently, the cases of the fermionic and gauge fields were shown to be regular distributions and treated consistently. In particular, we obtained the retarded distributions and determined the causal and Feynman propagators in the proper light-front contour. To conclude our analysis, we dealt with the general light-front singularity of the type $g\left(k;n\right)=\left( k^{+}\right) ^{-n}$ in the framework of the causal method. By following the same steps as the ones presented to the fermionic and gauge fields cases, we were able to determine the general and well-defined expression to the Feynman propagator. In two particular cases, for a suitable choice for the $k_\lambda$ \textit{parameter}, we were able to reproduced the well-known results of the Mandelstam-Leibbrandt and Pimentel-Suzuki prescriptions.

We believe to have in hands all the necessary tools and results to perform more realistic analysis. With the present results we can make use of the framework of Epstein-Glaser causal method and study interacting theories in the light-front,
such as the Quantum Electrodynamics and Quantum Chromodynamics. There are still dubious problems present in previous analysis in the literature and we believe that the causal analysis be the proper way to obtain well-defined and unambiguous outcomes. These issues and others will be further elaborated, investigated and reported elsewhere.

\subsection*{Acknowledgments}

The authors would like to thanks the anonymous referee for his/her comments and suggestions to improve this paper.
R.B. thanks FAPESP for full support, B.M.P. thanks CNPq and CAPES for partial support and D.E.S. thanks CNPq for full support.

\appendix


\section{Light-front and general notations}

\label{sec:appA}

If the set of points $\left( x^{0},x^{1},x^{2},x^{3}\right) $ are the usual instant-form
coordinate system, one may introduce the standard relations%
\begin{align*}
\hat{x}^{0,3} &= \frac{x^{0}\pm x^{3}}{\sqrt{2}}\equiv x^{\pm} ,\\
\hat{x}^{\bot} &= \left(x^{1}, x^{2} \right),
\end{align*}
as the set of points $\left( x^{+},x^{1},x^{2},x^{-}\right) $ in the light-front coordinate system. Moreover, the metric
in the light-front form is written
\begin{equation}
h_{\mu \nu }=h^{\mu \nu }=\allowbreak
\begin{pmatrix}
0 & 0 & 0 & 1 \\
0 & -1 & 0 & 0 \\
0 & 0 & -1 & 0 \\
1 & 0 & 0 & 0%
\end{pmatrix}. \label{A.0}
\end{equation}
The invariant inner product between four-vectors takes the form
\begin{equation}
A_{\mu}A^{\mu} = 2A_{+}A_{-}-A_{i}A_{i},
\end{equation}
where the components of a vector $A^{\mu }=\left(A^{+},A^{\bot},A^{-}\right)$ are usually denoted as the temporal, transversal and longitudinal components, respectively. Whenever convenient we also use the notation $ \hat{A} $ for the transverse part.

Throughout the paper we have used the following definition for the general Fourier and the inverse Fourier transforms
\begin{align}
\hat{d}\left(p\right)&=\left(2\pi\right)^{-m/2}\int d^m x d\left(x\right)e^{ip.x},\label{A.1}\\
\check{d}\left(p\right)&=\left(2\pi\right)^{-m/2}\int d^m x d\left(x\right)e^{-ip.x}.\label{A.2}
\end{align}
respectively, with $m$ the spacetime dimension.


\section{Fourier transform for an inverse-differential operator}

Let us consider $P\left( x,\partial \right) $ a differential operator and $d\left(x\right) $ a distribution, by simplicity, defined in one dimension. We can represent the operation $P\left( x,\partial \right) d\left( x\right) $ as the following
Fourier transformation%
\begin{equation}
P\left( x,\partial \right) d\left( x\right) =\left( 2\pi \right)
^{-1/2}\int\limits_{-\infty }^{+\infty }dkP\left( -ik\right) \hat{d}\left(
k\right) e^{-ikx},
\end{equation}%
where $\hat{d}$ is the Fourier transformed distribution of $d$. The quantity $P\left( -ik\right) $ is a polynomial for the differential operator. This result can be extended for any other kind of operators. Then, for an arbitrary operator $A$, it can be defined as follows
\begin{equation}
\left( Ad\right) \left( x\right) =\left( 2\pi \right)
^{-1/2}\int\limits_{-\infty }^{+\infty }dka\left( -ik\right) \hat{d}\left(
k\right) e^{-ikx},
\end{equation}%
where $a\left( -ik\right) $ is some function associated to the operator $ A $. Hence, this representation can be used to define inverse-differential operators. In particular, the inverse operator $ \left( \partial \right) ^{-n} $ can be defined as follows
\begin{equation}
\left( \partial \right) ^{-n}d\left( x\right) \equiv \left( 2\pi \right)
^{-1/2}P\int\limits_{-\infty }^{+\infty }dk\frac{1}{\left( -ik\right) ^{n}}%
\hat{d}\left( k\right) e^{-ikx}, \quad n=1,2,\ldots
\end{equation}%
where $P$ indicates the principal Cauchy value. We develop this idea for the $\delta $-Dirac distribution, in which 
$\delta \left( x\right) =\left( 2\pi \right) ^{-1}\int dk e^{-ikx}$. Then, $ \left( \partial \right) ^{-n}\delta \left( x\right) $ can be written in the form%
\begin{equation}
\left( \partial \right) ^{-n}\delta \left( x\right) =\left( 2\pi \right)
^{-1}\frac{1}{\left( -i\right) ^{n}}P\int\limits_{-\infty }^{+\infty }dk%
\frac{1}{k^{n}}e^{-ikx}
\end{equation}
Moreover, using the relation:%
\begin{equation}
P\left( \frac{1}{k^{n}}\right) =\frac{1}{2}\left[ \frac{1}{\left(
k+i0^{+}\right) ^{n}}+\frac{1}{\left( k-i0^{+}\right) ^{n}}\right],
\end{equation}
and, after some calculation, we arrive at
\begin{equation}
\left( \partial \right) ^{-n}\delta \left( x\right) =\left( 2\pi \right)
^{-1}\frac{1}{2}\frac{\left( -ix\right) ^{n-1}}{\left( -i\right) ^{n}\left(
n-1\right) !}\left[ \int\limits_{-\infty }^{+\infty }dk\frac{1}{\left(
k+i0^{+}\right) }e^{-ikx}+\int\limits_{-\infty }^{+\infty }dk\frac{1}{\left(
k-i0^{+}\right) }e^{-ikx}\right].
\end{equation}
On the other hand, identifying the terms within brackets with the Heaviside-function
\begin{equation}
\theta \left( \pm y\right) =\left( 2\pi \right) ^{-1}\int\limits_{-\infty
}^{+\infty }dk\frac{i}{\pm k+i0^{+}}e^{-iky},
\end{equation}
we finally find that
\begin{equation}
\left( \partial \right) ^{-n}\delta \left( x\right) =\frac{1}{2}sgn\left(
x\right) \frac{x^{n-1}}{\left( n-1\right) !}, \quad n=1,2,\ldots \label{b.1}
\end{equation}%
where $sgn\left( x\right)
=\theta \left( x\right) -\theta \left( -x\right) $ is the sign function.
\label{sec:appB}




\section*{References}

\begin{thebibliography}{99}
\bibitem{1} P.A.M. Dirac, Rev. Mod. Phys. \textbf{21}, 392 (1949); B.L.G. Bakker, \emph{Forms of Relativistic Dynamics}, in: H. Latal, W. Schweiger (Eds.),  Lecture Notes in Physics Vol. 572,  Springer, New York, 2001, pp. 1-54.

\bibitem{34} K. Sundermeyer, \emph{Constrained Dynamics}, Lecture Notes in Physics Vol. 169, (Springer, New York, 1982).

\bibitem{35} T. Heinzl, \emph{Light-Cone Quantization: Foundations and Applications}, in: H. Latal, W. Schweiger (Eds.), Lecture Notes in Physics Vol. 572, Springer, New York, 2001, pp. 55-142.

\bibitem{30} S.J. Brodsky, H.C. Pauli and S.S. Pinsky, Phys. Rep. \textbf{301}, 299 (1998).

\bibitem{37} S.J. Brodsky, arXiv:hep-ph/9710288v2; S.J. Brodsky and H.C. Pauli, \emph{Light-cone quantization of quantum chromodynamics}, in: H. Mitter, H. Gausterer (Eds.), Lecture Notes in Physics Vol. 396, Springer, Berlin, 1991, pp. 51-121; S.J. Brodsky and G.P. Lepage,
            \emph{Perturbative Quantum Chromodynamics}, World Scientific, Singapore, 1989; C. B. Thorn, Phys. Rev. D
            \textbf{20}, 1435 (1979); Phys. Rev. D \textbf{20}, 1934 (1979); R.J. Perry, arXiv:nucl-th/9901080v1; B.L.G. Bakker et al, arXiv:hep-ph/1309.6333v1.

\bibitem{31} J.B. Kogut and D.E. Soper, Phys. Rev. D \textbf{1}, 2901 (1970).

\bibitem{40} R.A. Neville and F. Rohrlich, Nuovo Cimento A \textbf{%
1}, 625 (1971); F. Rohrlich, Acta Phys. Austriaca, Suppl. VIII, 277 (1971); R. Casana, B.M. Pimentel, and
G.E.R. Zambrano, arXiv:hep-th/0803.2677; G.E.R. Zambrano, \emph{Formula\c{c}\~{a}o Can\^{o}nica no Plano Nulo}, Ph.D. Thesis, IFT-T.001/09.

\bibitem{41} S.J. Brodsky, R. Roskies and R. Suaya, Phys. Rev. D \textbf{8}, 4574 (1973); J.H. Ten Eyck and F. Rohrlich, Phys. Rev. D \textbf{9}, 2237 (1974).

\bibitem{9} E. Tomboulis, Phys. Rev. D \textbf{8}, 2736 (1973); A.T. Susuki and J.H.O. Sales, Mod. Phys. Lett. A \textbf{19},
1925 (2004); B.M. Pimentel, A.T. Suzuki and G.E.R. Zambrano, Few-Body Syst.
\textbf{52}, 437 (2012).

\bibitem{32} D.J. Pritchard and W.J. Stirling, Nucl. Phys. B \textbf{165}, 237 (1980).

\bibitem{33} G. Cursi, W. Furmanski and R. Petronzio, Nucl. Phys. B \textbf{175}, 27 (1980).

\bibitem{2} J. Schwinger, Phys. Rev. \textbf{130}, 402 (1963).

\bibitem{3} D.M. Capper and G. Leibbrandt, Phys. Rev. D \textbf{25}, 1002
(1982); \textbf{25}, 1009 (1982).

\bibitem{4} S. Mandelstam, Nucl. Phys. B \textbf{213}, 149 (1983).

\bibitem{5} G. Leibbrandt, Phys. Rev. D \textbf{29}, 1699 (1984).

\bibitem{7} B.M. Pimentel and A.T. Suzuki, Phys. Rev. D \textbf{42}, 2115
(1990); Mod. Phys. Lett. A \textbf{6}, 2649 (1991).

\bibitem{8} B.M. Pimentel, A.T. Suzuki and J.L. Tomazelli, Nuovo Cimento
\textbf{111}, 751 (1996).

\bibitem{10} A.S. Wightman, Phys. Rev. \textbf{101}, 860 (1956); R.F. Streater and A.S. Wightman, \emph{PCT, Spin and Statistics, and all that%
}, Princeton University Press, New Jersey, 2000; F. Strocchi, \emph{Selected Topics on the General Properties of Quantum Field Theory}, Lecture Notes in Physics Vol. 51, World Scientific, Singapore, 1993.

\bibitem{11} J.D. Bjorken and S.D. Drell, \emph{Relativistic Quantum Fields},
MacGraw-Hill, New York, 1965.

\bibitem{12} N.N. Bogoliubov and D.V. Shirkov, \emph{Introduction to the
Theory of Quantized Fields}, 3rd ed., John Wiley $\& $ Sons, New York, 1980.

\bibitem{13} P. Roman, \emph{Introduction to Quantum Field Theory}, John
Wiley $\& $ Sons, New York, 1969.

\bibitem{14} G. K\"{a}ll\'{e}n, \emph{Quantum Electrodynamics}, Springer-Verlag, New York, 1973.

\bibitem{6} A. Bassetto, G. Nardelli and R. Soldati, \emph{Yang-Mills Theories in Algebraic non-Covariant Gaugess}, World Scientific,
Singapore, 1991; G. Leibbrandt, \emph{Noncovariant Gauges: Quantization of
Yang-Mills and Chern-Simons Theory in Axial-Type Gauges}, World Scientific, Singapore, 1994.

\bibitem{15} R.D. Carmichael and D. Mitrovic, \emph{Distribution and
analytic functions}, John Wiley $\& $ Sons, New York, 1989.

\bibitem{16} V.S. Vladimirov, \emph{Methods of the theory of Generalized functions}, Taylor \& Francis, London, 2002; A.S. Demidov, \emph{Generalized Functions in Mathematical Physics}, 2nd ed., Nova Science Publishers, 2013.

\bibitem{18} H. Epstein and V. Glaser, Ann. Inst. H. Poincar\'{e} A \textbf{%
19}, 211 (1973).

\bibitem{19} G. Scharf, \emph{Finite Quantum Electrodynamics: The Causal
Approach}, 2nd ed., Springer-Verlag, Berlin, 1995.

\bibitem{20} A. Aste, Ann. Phys. \textbf{257}, 158 (1997); PoS 001 (LC2008) arXiv:hep-th/0810.2173v1.

\bibitem{36} A. Das, J. Frenkel and Silvana Perez, Phys.Rev. D \textbf{70}, 125001 (2004); A. Das and J. Frenkel, Phys.Rev. D \textbf{71}, 087701 (2005).

\bibitem{22} P. Grang\'{e} and E. Werner, Nucl. Phys. Proc. Suppl. B \textbf{%
161}, 75 (2006); B. Mutet, P. Grang\'{e} and E. Werner, PoS 005 (LC2008).

\bibitem{23} P. Grang\'{e} and E. Werner, arXiv:math-ph/0612011; Nucl. Phys.
Proc. Suppl. B \textbf{161}, 75 (2006).

\bibitem{24} P. Grang\'{e}, J.-F. Mathiot, B. Mutet and E. Werner, Phys.Rev.
D \textbf{80}, 105012 (2009); Nucl. Phys. B, Proc. Suppl. \textbf{199}, 191
(2010); Phys.Rev. D \textbf{82}, 025012 (2010).

\bibitem{17} G.C. Wick, Phys. Rev. \textbf{96}, 1124 (1954).

\bibitem{49} R. Bufalo, B.M. Pimentel and D.E. Soto, Ann. Phys. \textbf{351}, 1062 (2014).

\bibitem{21} J. Hilgevoord, \emph{Dispersion Relations and Causal Description%
}, North-Holland Publishing, Amsterdam, 1960.

\bibitem{bp} N.N. Bogoliubov and O.A. Parasiuk, Acta Math. 97, 227 (1957).

\bibitem{37a} D.M. Capper, J.J. Dulwich and M.J. Litvak, Nucl. Phys. B \textbf{241}, 463
(1984); R. Bent\'in and A.T. Suzuki, Mod. Phys. Lett. A \textbf{22}, 1329 (2007).



\end{thebibliography}
\end{document}